\documentclass[fleqn,10pt]{wlscirep}
\usepackage[utf8]{inputenc}
\usepackage[T1]{fontenc}
\usepackage{siunitx}
\mathchardef\mhyphen="2D
\usepackage{hyperref}

\usepackage{multirow}
\usepackage[flushleft]{threeparttable} %
\usepackage{array}
\usepackage{upgreek}
\usepackage{eucal}
\usepackage{amsmath}

\newcommand{\eqnref}[1]{Eq.~\eqref{Eq:#1}}

\newcommand{\Suppltabref}[1]{Supplementary~Table~\ref{tab:#1}}

\newcommand{\figref}[1]{Fig.~\ref{fig:#1}}
\newcommand{\Supplfigref}[1]{Supplementary~Figure~\ref{fig:#1}}
\newcommand{\figrefL}[1]{Figure~\ref{fig:#1}}
\newcommand{\subfigref}[2]{Fig.~\hyperref[fig:#1]{\ref*{fig:#1}#2}}
\newcommand{\Supplsubfigref}[2]{Supplementary~Figure~\hyperref[fig:#1]{\ref*{fig:#1}#2}}
\newcommand{\subfigrefL}[2]{Figure~\hyperref[fig:#1]{\ref*{fig:#1}#2}}
\newcommand{\subfigsref}[3]{Figs.~\hyperref[fig:#1]{\ref*{fig:#1}#2}-\hyperref[fig:#1]{\ref*{fig:#1}#3}}
\newcommand{\Supplsubfigsref}[3]{Supplementary~Figs.~\hyperref[fig:#1]{\ref*{fig:#1}#2}-\hyperref[fig:#1]{\ref*{fig:#1}#3}}
\newcommand{\subfigsrefL}[3]{Figures~\hyperref[fig:#1]{\ref*{fig:#1}#2}-\hyperref[fig:#1]{\ref*{fig:#1}#3}}
\newcommand*{\sspds}[0]{SNSPDs}
\newcommand*{\sspd}[0]{SNSPD}

\usepackage{placeins}

\title{Bright Quantum Dot Single-Photon Emitters at Telecom Bands Heterogeneously Integrated on Si}
\author[1,2,*]{P.~Holewa}
\author[2]{A.~Sakanas}
\author[3]{U.~M.~G\"{u}r}
\author[1]{P.~Mrowi\'nski}
\author[4]{A.~Huck}
\author[5,2]{B.~Wang}
\author[1]{A.~Musia\l{}} 
\author[2,6]{K.~Yvind}
\author[2]{N.~Gregersen}
\author[1,+]{M.~Syperek} 
\author[2,6,++]{E.~Semenova} 

\affil[1]{Laboratory for Optical Spectroscopy of Nanostructures, Faculty of Fundamental Problems of Technology, Department of Experimental Physics, Wroc\l{}aw University of Science and Technology, Wyb. Wyspia\'{n}skiego 27, 50-370 Wroc\l{}aw, Poland}

\affil[2]{DTU Fotonik, Technical University of Denmark, Kongens Lyngby 2800, Denmark}

\affil[3]{DTU Electrical Engineering, Technical University of Denmark, 2800 Kongens Lyngby, Denmark}

\affil[4]{Center for Macroscopic Quantum States (bigQ), Department of Physics, Technical University of Denmark, 2800 Kongens Lyngby, Denmark}

\affil[5]{Hefei National Laboratory for Physical
Sciences at Microscale, University of Science and Technology of China, Hefei, Anhui 230026, China}

\affil[6]{NanoPhoton-Center for Nanophotonics, Technical University of Denmark, 2800 Kongens Lyngby, Denmark}

\affil[*]{pawel.holewa@pwr.edu.pl}

\affil[+]{marcin.syperek@pwr.edu.pl}

\affil[++]{esem@fotonik.dtu.dk}

\begin{abstract}
Whereas the Si photonic platform is highly attractive for scalable optical quantum information processing, it lacks practical solutions for efficient photon generation.
Self-assembled semiconductor quantum dots (QDs) efficiently emitting photons in the telecom bands ($\SIrange{1460}{1625}{\nano\meter}$) allow for heterogeneous integration with Si.
In this work, we report on a novel, robust, and industry-compatible approach for achieving single-photon emission from InAs/InP QDs heterogeneously integrated with a Si substrate.
As a proof of concept, we demonstrate a simple vertical emitting device, employing a metallic mirror beneath the QD emitter, and experimentally obtained photon extraction efficiencies of $\sim10\%$. Nevertheless, the figures of merit of our structures are comparable with values previously only achieved for QDs emitting at shorter wavelength or by applying technically demanding fabrication processes.
Our architecture and the simple fabrication procedure allows for the demonstration of a single-photon generation with purity $\mathcal{P}>98\%$ at the liquid helium temperature and $\mathcal{P}=75\%$ at $\SI{80}{\kelvin}$.
\end{abstract}
\begin{document}

\flushbottom
\maketitle

\thispagestyle{empty}

\section*{Background}
Exploiting single photons as a resource is a powerful approach for quantum information processing (QIP)~\cite{Knill2001,Kok2007,Wang2019Boson,Elshaari2020}. Photons have long coherences and efficiently propagate over macroscopic distances, which enabled the demonstration of a computational advantage with a quantum photonic processor~\cite{Zhong2020}, loophole-free tests of Bell's theorem~\cite{Shalm2015,Giustina2015}, and long distance quantum key distribution~\cite{Takemoto2015,Yin2020}. Scalability of optical QIP requires the miniaturization, coupling and integration of active and passive photonic components into quantum photonic integrated circuits (QPICs)~\cite{Wang2019NatPhot}. The Si-based photonic platform is a leading candidate for integration with transparency wavelengths $>\SI{1.1}{\micro\meter}$ and mature manufacturing processes~\cite{Wang2018, Elshaari2020}. QPICs supporting multi-dimensional entanglement~\cite{Wang2018} and quantum processors with hundreds of elements~\cite{Qiang2018, Wang2018} have been demonstrated. Si however does not allow for efficient light generation. Spontaneous four-wave mixing is the commonly employed approach for single photon generation on Si and realized with Si-on-insulator integration~\cite{Wang2018,Qiang2018,Paesani2019}. This process is probabilistic with few-percent efficiency~\cite{Wang2019NatPhot}, thus limiting scalability. Hybrid approaches combining solid-state single photon emitters (SPEs) with Si have instead been investigated~\cite{Kim2017Emitter}, but so far requiring technically demanding fabrication. The reliable realization of SPEs monolithically integrated with Si and allowing for the deterministic emission of pure single photons remains challenging.

Among different candidates~\cite{Eisaman2011,Aharonovich2016}, epitaxially grown self-assembled semiconductor quantum dots (QDs)~\cite{Senellart2017} emitting in the long-wavelength telecom bands ($\SIrange{1460}{1625}{\nano\meter}$)~\cite{Arakawa2020} are suitable for integration with Si~\cite{Tran2018,Wang2018,Vlasov2004}. The telecom wavelength promises very low Si-waveguide propagation losses~\cite{Tran2018} and allows for interconnecting distant QIPCs using optical fiber networks and for distributed quantum computing~\cite{Wehner2018,Cuomo2020}. Photon emission in the telecom bands has been achieved with QDs based on either InAs/GaAs~\cite{Semenova2008,Paul2017,Zeuner2019,Nawrath2019,Musial2020} or InAs/InP~\cite{Takemoto2007,Miyazawa2016,Birowosuto2012,Benyoucef2013,Skiba-Szymanska2017,Muller2018,Anderson2020,Anderson2020PRApplied,Holewa2020PRB} material composition, and excellent quantum light sources with high purity and indistinguishability~\cite{Cao2019,Arakawa2020} were demonstrated, fulfilling the requirements for QIP~\cite{Senellart2017}.

The photon extraction efficiency $\eta$ for as-grown QD-based SPEs is typically $<1\%$~\cite{Senellart2017} due to the large semiconductor-air refractive index contrast, but can be increased by tailoring the local optical environment~\cite{Claudon2010,Stepanov2015,Liu2019}. Common approaches for increasing $\eta$ include placing a QD in a monolithic cavity defined by distributed Bragg reflectors (DBRs)~\cite{Muller2018}, in an optical horn structure~\cite{Takemoto2007}, or atop a single DBR reflector~\cite{Benyoucef2013, Paul2017, Zeuner2019, Musial2021}. DBR-based approaches are scalable and $\eta$ up to $13\%$ was achieved in a narrow spectral window~\cite{Musial2021}, following demanding fabrication in the InAs/InP material system due to layers' low refractive index contrast.
With the horn structure, $\eta\approx11\%$ at $\SI{1560}{\nano\meter}$ (numerical aperture NA$=0.55$) was shown~\cite{Takemoto2007}, requiring complex fabrication. These approaches however are not suitable for the monolithic integration of QD-based SPEs with Si. 

In this work, we propose and demonstrate efficient single photon emission with $\eta>10\%$ (NA$=0.4$) and wavelength in the telecom bands. Our photon sources are based on InAs QDs epitaxially grown on InP and heterogeneously integrated on a Si substrate via chemical bonding. We achieve triggered photon emission with a generation purity of $>98\%$ at a temperature of $T=\SI{4.2}{\kelvin}$ and $\sim75\%$ at $\SI{80}{\kelvin}$. Further increase of $\eta$ is possible employing the higher NA objective and tailoring the QD optical environment. Our approach promises localizing individual QDs via optical imaging~\cite{Sapienza2015} and subsequent processing of photonic components with deterministic spatial alignment. Moreover, our QD integration method on the Si-platform provides a broad range of device architecture possibilities, in particular the in-plane emission into planar waveguides as required for on-chip integration.

\section*{Results}

	\begin{figure*}[!tb] %
		\begin{center} %
    	\includegraphics[width=1\columnwidth]{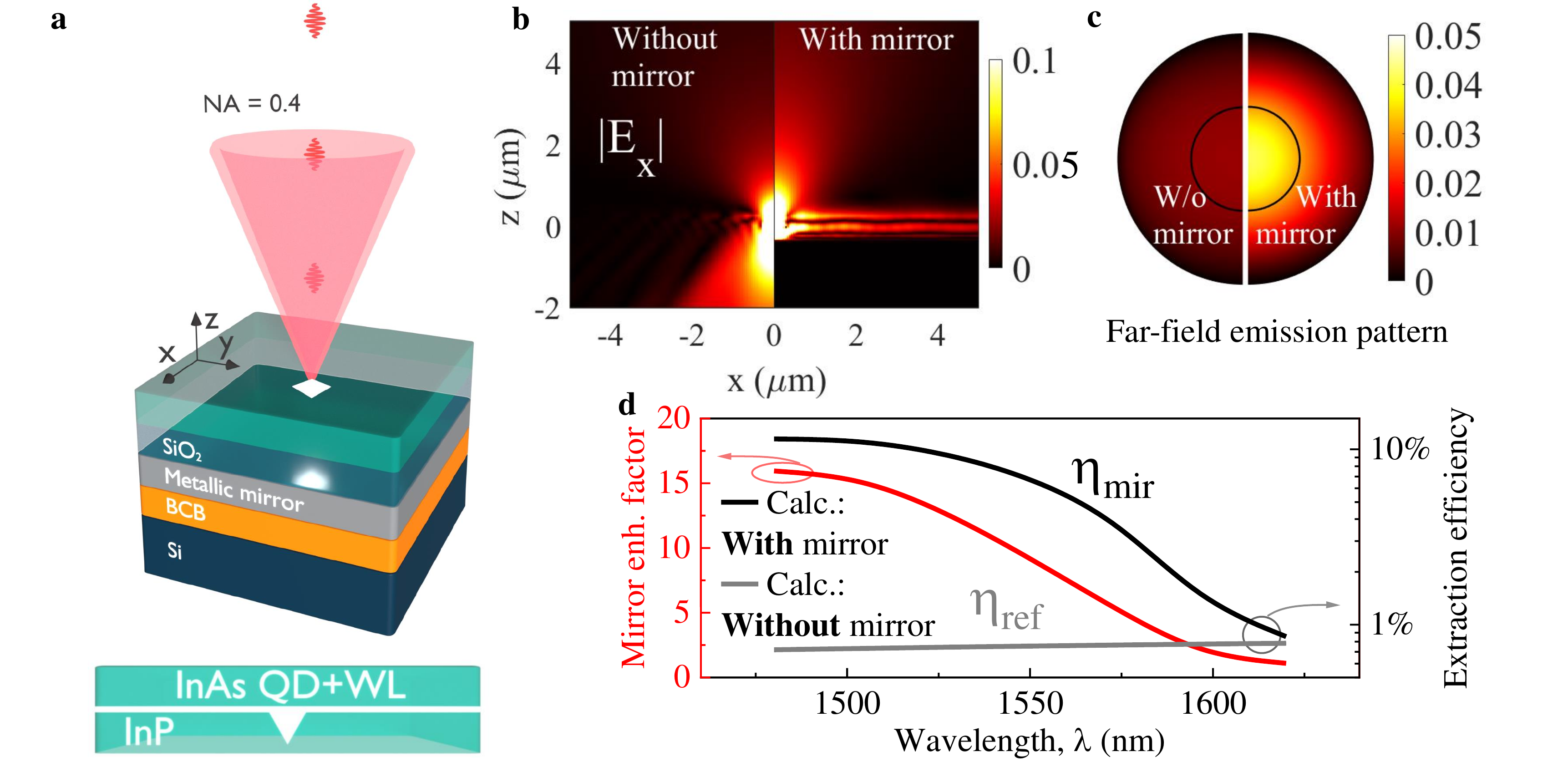} 
	    \end{center}	 %
		\caption{\label{fig:Structure-layout}
		The design of our structures and theoretically estimated performance.
		\textbf{a}, The investigated structure scheme, consisting of InAs/InP quantum dots (QDs) with a metallic reflector integrated on a Si substrate. WL -- wetting layer.
        \textbf{b}, The electric field component $|E_{\textrm{x}}|$ for $\lambda=\SI{1550}{\nano\meter}$ for the structure without (left) and with (right) a metallic reflector made of aluminum (Al).
        \textbf{c}, The calculated far field emission ($P_{\textrm{lens, NA}}$) for the reference (left) and the device with Al-mirror (right). The half-circle marks the collection cone of a 0.4~NA objective.
        \textbf{d}, The calculated broadband mirror enhancement factor (left axis) and photon extraction efficiency for the QD device with mirror ($\eta_{\textrm{mir}}$) and the reference structure without mirror ($\eta_{\textrm{ref}}$) as function of emitter wavelength.}
	\end{figure*}
\subsection*{Structure design}
The schematic design of our structures together with calculated device performances are presented in~\subfigref{Structure-layout}. Self-assembled InAs QDs are placed in a weak planar cavity system formed between a bottom metallic mirror and a top InP/air interface, as illustrated in ~\subfigref{Structure-layout}{a}. We determine the achievable photon extraction efficiency using numerical simulations of the electromagnetic (EM) field (\subfigref{Structure-layout}{a}) generated by the InAs QD modelled as a classical point dipole with in-plane dipole orientation.
The EM field distribution ($|E_{\textrm{x}}|$) is presented in~\subfigref{Structure-layout}{b} for a reference structure without (left panel) and with a metal mirror (right panel). We observe that the effect of the mirror is to suppress leakage of light into the substrate and instead direct light in the vertical direction. While additional in-plane guiding of light in the slab waveguide formed by the air-semiconductor-metal interfaces is visible, the out-of-plane field pattern is significantly enhanced compared to the structure without reflector. Furthermore, the enhancement is observed in the far field emission pattern presented in~\subfigref{Structure-layout}{c}, highlighting the role of the metallic mirror in the directional emission. The black circle represents the collection aperture of a typical, long-working distance microscope objective with a 0.4 NA used for light collection in the experiment, and the extraction efficiency $\eta$ is then defined as the ratio of the power collected within the NA of the objective ($P_{\textrm{lens, NA}}$) to the total power emitted by the dipole (\hyperref[sec:Extraction-efficiency-calc]{Methods. Numerical calculations.}). The computed extraction efficiency is presented in \subfigref{Structure-layout}{d} as a function of wavelength. We define the mirror enhancement factor as the ratio between the extraction efficiency for the planar structure with mirror ($\eta_{\textrm{mir}}$) and the reference structure ($\eta_{\textrm{ref}}$).
The presence of the metallic reflector leads to a broadband enhancement of the extraction efficiency (left axis), with the 9.2-fold increase at $\SI{1550}{\nano\meter}$ and $\eta\approx7\%$, and nearly 16-fold increase at $\SI{1500}{\nano\meter}$ and $\eta\approx11\%$. This expected performance is competitive with state-of-the-art extraction efficiencies~\cite{Takemoto2007,Musial2021} for single-photon sources operating in the long-wavelength telecom bands.

The device fabrication begins with the epitaxy of an InGaAs sacrificial layer lattice-matched to a standard (001) oriented InP substrate, followed by the growth of an InP $\lambda$--cavity with an array of low surface density ($\SI{\sim2.8e9}{\centi\meter\tothe{-2}}$) InAs QDs placed in the center for quantum confinement  (\hyperref[sec:QDs-fabrication]{Methods. QD fabrication.}). In the next step, the top InP surface is covered by $\SI{100}{~\nano\meter}$ of SiO$_2$ followed by a $\SI{100}{~\nano\meter}$-thick metallic reflector (Al in our case). Subsequently, the chip is flipped and bonded to a Si substrate using benzocyclobutene (BCB) and finally, the thick InP substrate, now on top, and the InGaAs sacrificial layer are removed (\hyperref[sec:Integration]{Methods. QD integration with Si substrate.}). We emphasize that this approach and the dimensions are suitable for in-plane photon emission into a Si photonic circuit, although not explicitly pursued in this work.
	\begin{figure*}[!tb] %
		\begin{center} %
			\includegraphics[width=1\columnwidth]{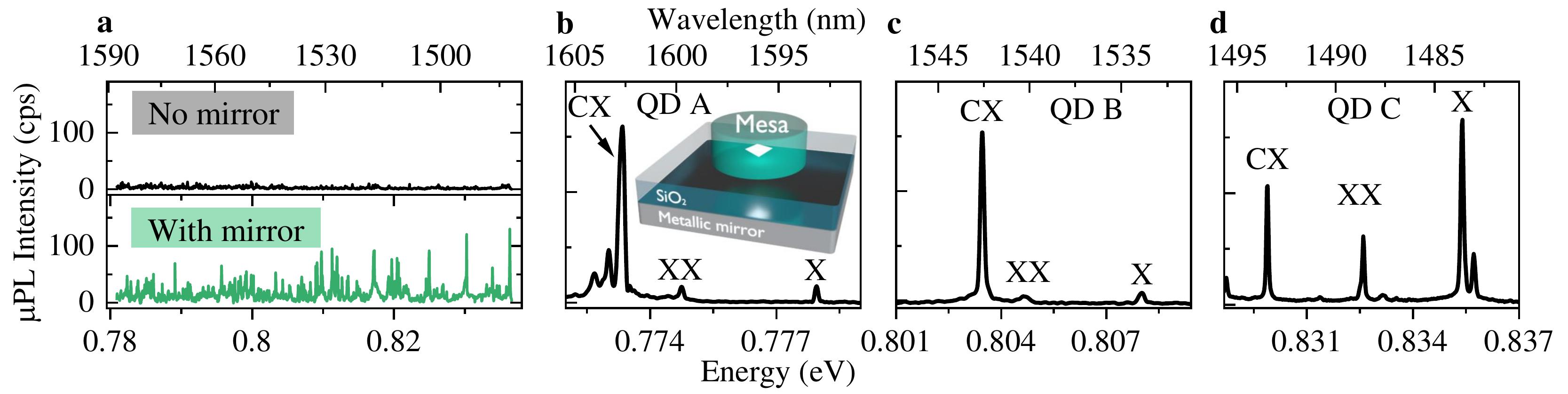} %
		\end{center} %
		\caption{\label{fig:uPL spectra} Excitonic complexes in InAs/InP QDs.
    \textbf{a},~Representative high-spatially-resolved photoluminescence ($\upmu$PL) spectra recorded for the reference structure without mirror (top panel) and the structure with mirror (bottom panel) with identical pulsed laser excitation at $T=\SI{4.2}{\kelvin} $.
    \textbf{b}-\textbf{d},~$\upmu$PL spectra of the investigated InAs/InP QDs (labelled A, B, and C) with identified excitonic emission complexes: neutral exciton (X), biexciton (XX), and charged exciton (CX). Inset to \textbf{b}: mesa structure.}
	\end{figure*}
We applied a two-fold experimental evaluation strategy to verify the significant robustness of the structure design with respect to the level of $\eta$ and the related broadband performance, and to present bright SPEs heterogeneously integrated on a Si substrate.
In \subfigrefL{uPL spectra}{a}, we present emission spectra obtained from a reference structure without metallic reflector and the planar structure containing the metallic mirror, both recorded with the same high spatial resolution photoluminescence setup ($\upmu$PL) from a diffraction limited spot (\hyperref[sec:OptExp]{Methods. Optical experiments.}). The spectra consist of multiple sharp emission lines distributed over a broad spectral range, originating from QDs of mainly different sizes within the optical excitation spot and various emission complexes including neutral exciton (X), charged exciton (CX), and biexciton (XX) transitions. For the structure containing the mirror and compared to the reference structure, we observe a clear intensity enhancement of the emission lines.

The quantitative analysis of photon extraction efficiency $\eta$ from SPEs requires the isolation of single QDs and identification of their respective spectral features.
We therefore proceeded with the processing of the mirror-containing planar structure to fabricate cylindrical mesas with diameters of $\textrm{D}_1=\SI{2}{\micro\meter}$ and $\textrm{D}_2=\SI{3}{\micro\meter}$, respectively, as schematically illustrated in the inset of \subfigref{uPL spectra}{b}. The finite size of the mesas allows for the spatial isolation of single QDs, a vital element in single-photon source engineering, and leads to modifications of the calculated EM field pattern and extraction efficiencies (see \Supplfigref{FDTD-calculations}).

In \subfigsrefL{uPL spectra}{b}{d}, we present the $\upmu$PL spectra of three representative and isolated QD emitters, which in the following we refer to as QD A, B, and C, with their emission spectrum located in the telecom L-, C-, and S-band, respectively. The indicated excitonic complexes (X, CX, XX) are identified based on a series of excitation-power--dependent and polarization--resolved $\upmu$PL measurements (see \Supplfigref{Identification-of-lines}), and confirmed by the cross-correlation of the XX-X and CX-X complexes. (see \Supplfigref{Suppl-Cross-correlation}).
	\begin{figure}[!tb] %
		\begin{center} %
			\includegraphics[width=\columnwidth/2]{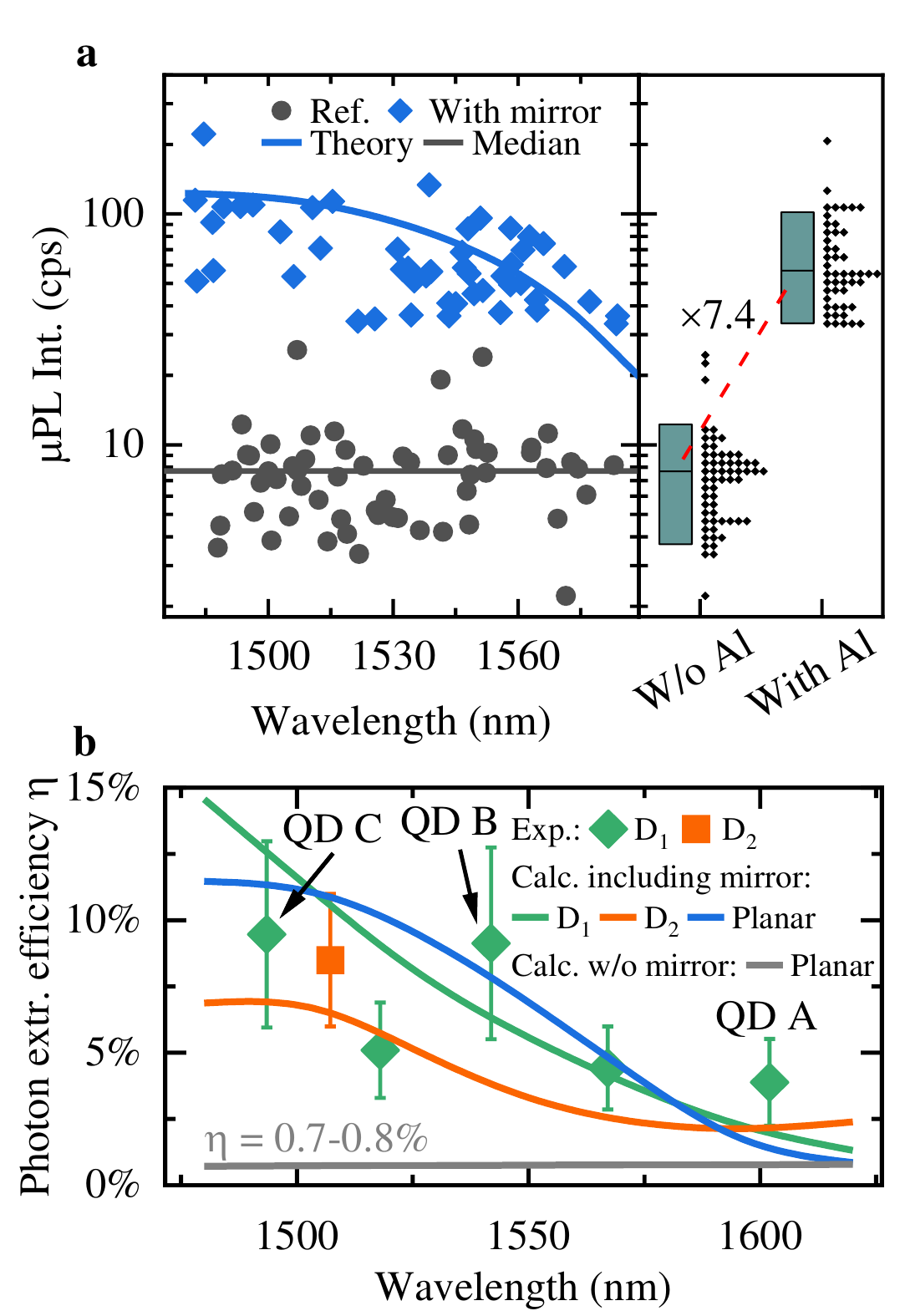} %
		\end{center} %
		\vspace{-0.8em}\caption{\label{fig:Extraction-efficiency}%
			Photon extraction efficiency for the investigated structures.
			\textbf{a},~Left panel: comparison of the $\upmu$PL intensity of $\sim50$ brightest emission lines (points) for the planar mirror-containing (blue diamonds) and the reference (black circles) structures, respectively.
			The solid blue line is the expected $\upmu$PL intensity for mirror-containing structure obtained by multiplying the median $\upmu$PL intensity of the reference structure (solid black line) by the mirror enhancement factor (cf.~red line in~\subfigref{Structure-layout}{d}).
			Right panel: statistical analysis of measured intensities. Boxes illustrate one standard deviation, the line inside the box is the median value of each distribution shown as points.
			\textbf{b},~Photon extraction efficiency $\eta$ for the mesa-processed structure with a metallic mirror. Green diamonds show recorded $\eta$ values for mesas with $\textrm{D}_1=\SI{2}{\micro\meter}$ including QDs A-C.
			The result shown as orange square is obtained for an emitter in a mesa with $\textrm{D}_2=\SI{3}{\micro\meter}$.
			The solid lines represent calculated $\eta$ values obtained with the modal method for mesas with $\textrm{D}_1=\SI{2}{\micro\meter}$ (green) and $\textrm{D}_2=\SI{3}{\micro\meter}$ (orange).
			Solid blue and grey lines show the calculated $\eta$ for a planar structure with mirror and a reference structure without mirror, respectively.}
	\end{figure}
\subsection*{Brightness of SPEs}
We employ two approaches to compare the calculated with the experimentally obtained photon extraction efficiency~$\eta$. In the first approach, we consider the broadband enhancement of the photon extraction for planar structures due to the mirror. The approach is based on the statistical analysis of correlated and uncorrelated emission processes, from where we derive a rough estimate of $\eta$. We thus compare the intensity of nearly 50 spectral lines recorded from the mirror-containing and the reference structure, respectively, and the results are plotted in \subfigref{Extraction-efficiency}{a}. The spectrally averaged enhancement factor of $7.4^{+1.6}_{-1.3}$ is obtained by comparing the median values for the distributions of emission intensity, which can be converted to a mean photon extraction efficiency $\eta\approx6\%$ for the device containing the metallic mirror. The 95\% confidence levels for the enhancement factor are calculated according to Ref.~\cite{Bonett2020}. In the second approach, we adopt the method described by M. Gschrey et al.~\cite{Gschrey2013} and directly measure $\eta$ for individual QDs in mesa-processed structures.
The results for $\eta$ from in total 6 QDs (including QDs A-C) are presented in \subfigref{Extraction-efficiency}{b} together with the numerically calculated values. For these mesa structures, we experimentally determine photon extraction efficiencies $\eta$ above $4.4\%$, and as high as $\eta_{\textrm{B}}=\SI{9.1}{\percent}$ and $\eta_{\textrm{C}} = \SI{9.5}{\percent}$ for QDs~B and~C, respectively. Those values demonstrate a one order of magnitude improvement compared to the efficiency $\SIrange{0.7}{0.8}{\percent}$ estimated for the reference structure without the metallic mirror. Importantly, we obtain good agreement between the experimentally determined extraction efficiency and the theoretical prediction, both for the broadband approach (\subfigref{Extraction-efficiency}{a}) and the individually investigated 6 QDs in mesas (\subfigref{Extraction-efficiency}{b}). We attribute the small deviations between simulation and experimental values to possible non-radiative recombination channels in the QD vicinity, changes in the mesa geometry due to fabrication imperfections and non-deterministic positioning of the QD within the mesa. 
While these effects generally result in a lower recorded photon flux compared to the theoretical prediction, we note that a $\sim \SI{200}{\nano\meter}$ displacement of a QD from the mesa center results in an increase in the extraction efficiency as compared to a QD placed in the center. Such a displacement may explain the high $\eta$ value obtained for QD B (see \Supplfigref{FDTD-calculations} for further details).

\subsection*{Evaluation of the photon purity}
\begin{figure}[!tb] %
		\begin{center} %
			\includegraphics[width=\columnwidth/2]{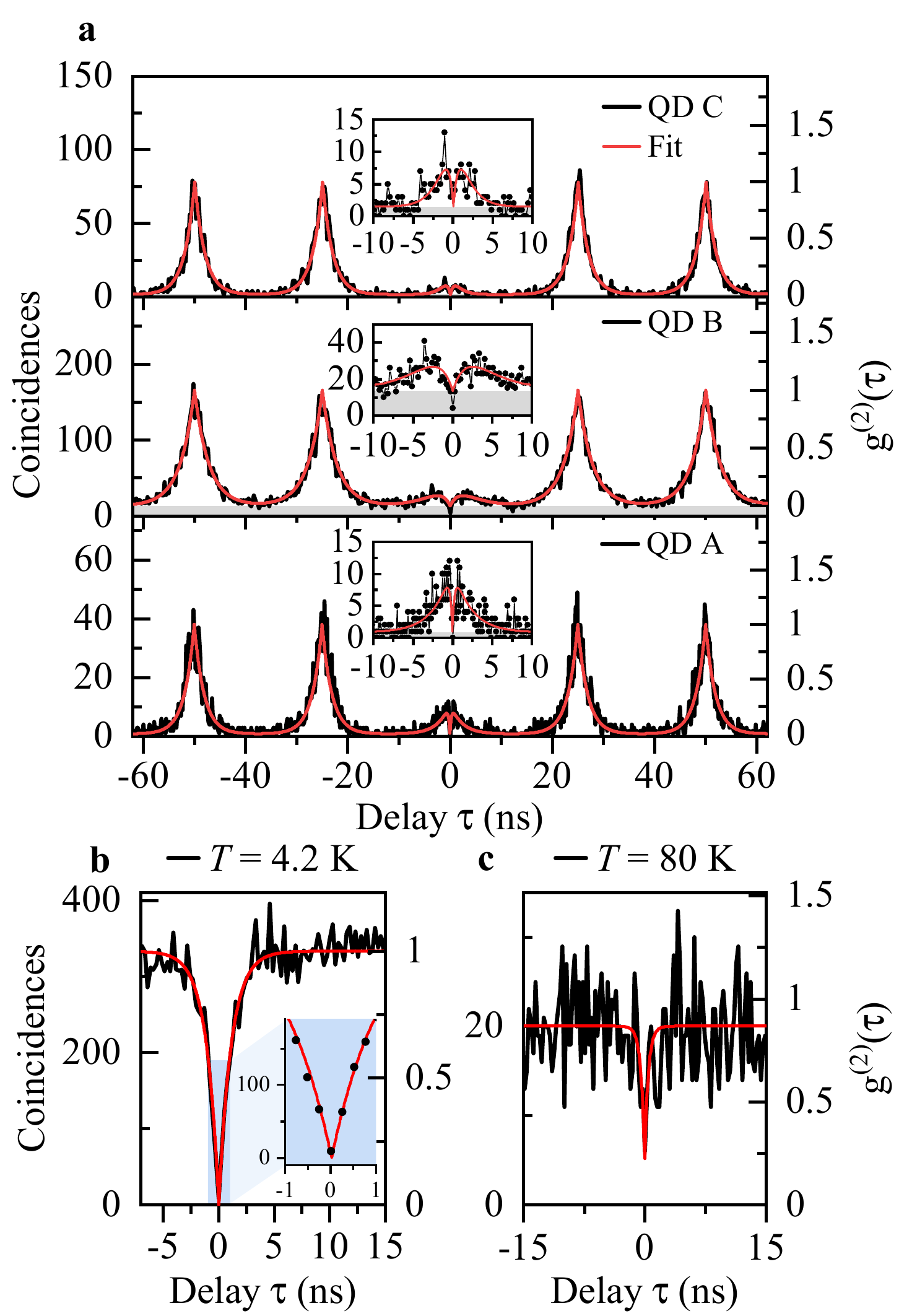} %
		\end{center} %
		\vspace{-0.8em}\caption{\label{fig:g2-pulsed}%
			Autocorrelation histograms for CX lines.
			\textbf{a},~Triggered single-photon emission for investigated QDs: C~(top), B~(center), A~(bottom). Insets:~Close-ups of the histograms showing coincidences around zero delay.
		    \textbf{b, c},~Single-photon emission under cw excitation for the CX in QD~B, recorded \textbf{b} under the laser excitation power corresponding to the saturation of the CX $\upmu$PL intensity (inset: zoom around $\tau = 0$), and \textbf{c} at $T=\SI{80}{\kelvin}$.
			Red lines are fits to the experimental data. Grey area in \textbf{a} shows the level of background coincidences $B$ obtained by the fit with \eqnref{g2-pulsed-Eq}.}
	\end{figure}

The order of magnitude improvement in photon extraction efficiency from InAs/InP QDs renders our structures an attractive source of single photons heterogeneously integrated with the Si-platform. We evaluate in the following the quality of single-photon emission from the QDs in mesa structures by investigating the purity of single photons emitted from QDs A-C. For that purpose, we recorded the second-order correlation function $g^{(2)}(\tau)$ exploiting off-resonant continuous wave (cw) and pulsed excitation schemes (\hyperref[sec:OptExp] {Methods. Optical experiments.}), and the obtained histograms without normalization are presented in \figrefL{g2-pulsed}. For pulsed excitation (\subfigref{g2-pulsed}{a}), we observe that the coincidences $\tau \approx 0 $ are strongly suppressed compared to the coincidence peaks at multiples of the inverse laser repetition rate ($\tau_0 = \SI{25}{\nano\second}$). Furthermore, we record a significant dip in the histogram counts at short time delays $|\tau|<\SI{5}{\nano\second}$ (insets to \subfigref{g2-pulsed}{a}). This feature indicates the capture of more than one carrier by the QD and cascaded photon emission within a single excitation, effectively resulting in multi-photon events~\cite{Fischbach2017} within $|\tau|<\SI{5}{\nano\second}$. We explain this observation with the off-resonant excitation scheme applied in our experiment, where a substantial amount of carriers are excited and trapped in the wetting layer or in other charge trap states~\cite{Peter2007,Kumano2016,Fischbach2017}. After release, those carriers are captured by the QD within the characteristic capture time $\tau_{\mathrm{cap}}$ and produce secondary photons~\cite{Chang2009} via exciton recombination with the time constant $\tau_{\mathrm{dec}}$.

We fit the correlation histograms obtained with pulsed laser excitation with the function~\cite{Dalgarno2008,Miyazawa2016}:
\begin{equation}\label{Eq:g2-pulsed-Eq}
C(\tau)=
B+A\left[\exp{\left(-|\tau|/\tau_{\mathrm{dec}}\right)}-\exp{\left(-|\tau|/\tau_{\mathrm{cap}}\right)}\right]+H \sum_{n\neq 0}\exp{\left(-|\tau-n\tau_0|/\tau_{\mathrm{dec}}\right)},
\end{equation}
where $B$ is the level of background coincidences, $A$ is a scaling parameter related to secondary photon emission, $n \neq 0$ the peak number and $H$ the average height of the peaks at $\tau_n = n \tau_0$. The second-order correlation function $g^{(2)}(\tau)$ is then obtained by normalizing $C(\tau)$ with $H$, and all fit parameters are summarized in \Suppltabref{g2-pulsed}.
Evaluating $g^{(2)}(\tau)$ at $\tau=0$, we obtain a single photon emission purity $\mathcal{P}=1-g^{(2)}(\tau=0)$ of $97.7 \pm 1.0~\%$, $91.3 \pm 1.7~\%$, and $98.2 \pm 1.2~\%$ for QDs A-C, respectively. This estimation ignores coincidences produced by secondary photon emission events which may be avoided asymptotically employing an on-resonance excitation scheme. Comparing the integrated peak area of $g^{(2)}(\tau=0)$ with the average peak area at $\tau_n$, as it is relevant for applications of our single photon sources in QIP, yields values for $\mathcal{P}_{\mathrm{area}}$ of 
$62.9 \pm 2.0~\%$, $56.7 \pm 1.8~\%$, and $79.5 \pm 2.0~\%$ for QDs A-C, respectively (see Methods).
We also correct the emission purity $\mathcal{P}_{\mathrm{area, cor}}$ for background coincidences $B$ and obtain vales of $72.4 \pm 2.0~\%$, $79.1 \pm 1.8~\%$, and $88.6 \pm 2.0~\%$ for QDs A-C. The emission purity recorded in the pulsed regime is mainly limited by the capture of secondary carriers and subsequent photon emission. We note that the data is well described by our model and does not consider the capture of secondary carriers at $\tau_n$.
Furthermore, with the modelling routine we determine $\tau_{\mathrm{dec}}$ in the range $\SIrange{1.9}{2.8}{\nano\second}$, which is in accordance with the CX decay time measured in time-resolved $\upmu$PL (see \Supplfigref{TRPL}).

Importantly, the obtained single photon purity of the investigated structures is very robust for a wide range of temperatures and excitation powers. In \subfigsref{g2-pulsed}{b}{c} we present correlation histograms recorded in CW excitation mode from the CX line of QD~B at sample temperatures of $\SI{4}{\kelvin}$ and $\SI{80}{\kelvin}$, respectively, while optically pumping at the saturation power $P_{\textrm{sat}}$ (see \Supplfigref{g2-Supplemental} for correlation histograms taken at $T=\SI{30}{\kelvin}$, $T=\SI{50}{\kelvin}$, at $0.3\times$ $P_{\textrm{sat}}$, $0.7\times$ $P_{\textrm{sat}}$, and \Suppltabref{Summary-fitted-parameters-cw} for the obtained $g^{(2)}(0)$ values).
At $\SI{80}{\kelvin}$ sample temperature, which easily can be reached with a liquid nitrogen dewar or a cryogen-free Stirling cooler~\cite{Musial2020}, a significant feature at $\tau=0$ is visible in the histogram, quantifying the robustness of this source of single photons. We fitted the normalized histograms recorded in cw mode with a standard single-exponential function~\cite{Michler2000a} (see Methods) to extract the single photon emission purity from the measurements.  
The raw data estimated purity from the histogram recorded at $P_{\textrm{sat}}$ is $\mathcal{P}_{\mathrm{raw}}=(97.3 \pm 2.7\%)$ (see Methods) limited by the finite time resolution of our setup.
The best fit to our data thus suggests even higher values of photon purity ($\mathcal{P}=100\%$), with a standard error $\sigma=0.038$, and without background correction.
Such high-purity of the single-photon flux in the high-excitation power regime ($P>=P_{\textrm{sat}}$) has previously been observed only for InAs/GaAs QDs emitting at $\lambda=(\SI{910}{\nano\meter}-\SI{920}{\nano\meter})$~\cite{Claudon2010,Gschrey2015}.
In contrast to these sources, the structure investigated here demonstrates high brightness and close-to-ideal single-photon purity while emitting in the telecom C-band.
Increasing the temperature to $T=\SI{80}{\kelvin}$, leads to a reduction of the emission line visibility (see \Supplfigref{QD-B-Temp-analysis} for the analysis of temperature-dependent $\upmu$PL of QD~B), yet we  obtain a relatively high purity of single photon emission with the fitted value of $\mathcal{P}_{\SI{80}{\kelvin}}=0.75$  ($\sigma=0.19$, without background correction).
We note that the achieved purity at $T=\SI{80}{\kelvin}$ is higher than the record values of $g^{(2)}(0)=0.34$ and $g^{(2)}(0)=0.33$ previously reported at this temperature and at $\SI{1.55}{\micro\meter}$ emission wavelength achieved with InAs/InAlGaAs/InP quantum dashes~\cite{Dusanowski2016,Arakawa2020} and InAs/GaAs QDs grown with the metamorphic approach~\cite{Carmesin2018}, respectively. Importantly, both these previous approaches are obtained in different carrier confinement conditions, InAlGaAs or GaAs barriers respectively, and thus cannot be directly compared with the pure InAs/InP system investigated here.

\section*{Discussion}
The demonstrated design of the structure with InAs/InP QDs on a metallic mirror integrated on a Si substrate paves the way towards a simplified, small-footprint, cost-effective, and scalable manufacturing process of triggered single-photon emitters operating in the telecom S-, C-, and L-bands, suitable for Si-based on-chip photonic quantum information processing. The spectral range of the InAs/InP single-photon emitters investigated here eliminates the necessity for frequency conversion to the telecom bands, potentially allowing for the implementation of distributed schemes for information processing and computation using low-loss fiber-based optical networks. Combining the robust design of our structures and the manufacturing process compatible with the existing industry standards establishes single-photon sources with high photon extraction efficiency in the broader telecom spectral range, with performance properties comparable to the DBR based solutions but with significantly reduced fabrication related technological demands.

The presented robust architecture, offering spectrally broad high photon extraction efficiency, is beneficial for further processing steps tailoring photonic environment of a deterministically localized emitter. The emitter high brightness allows for its fast spatial positioning utilizing the emission imaging method successfully employed for short-wavelength ($<\SI{1000}{\nano\meter}$) QDs~\cite{Sapienza2015}. However, at telecom wavelengths, the imaging relies on a 2D state-of-the-art InGaAs-based matrices with yet poor efficiencies and high noise levels compared to Si-based arrays desired for shorter wavelengths. Therefore, the simplified architecture of a QD on a metal mirror can open the route towards fabrication of fully deterministic, scalable QPICs at telecom wavelengths.

\section*{Methods}

\subsection*{QD fabrication}\label{sec:QDs-fabrication}
The QDs are grown in the low-pressure MOVPE TurboDisc\textregistered\ reactor using arsine (AsH$_3$), phosphine (PH$_3$), tertiarybutylphosphine (TBP) and trimethylindium (TMIn) precursors with H$_2$ as a carrier gas. The growth sequence starts with the deposition of a  $\SI{0.5}{\micro\meter}$-thick InP buffer layer on a (001)-oriented InP substrate at $\SI{610}{\degreeCelsius}$ subsequently epitaxially covered by a $\SI{200}{\nano\meter}$-thick In$_{0.53}$Ga$_{0.47}$As sacrificial layer lattice-matched to InP and a $\SI{244}{\nano\meter}$-thick InP layer. Then, the temperature is decreased to $\SI{483}{\degreeCelsius}$, stabilized under TBP for $\SI{180}{\second}$ and AsH$_3$ for $\SI{27}{\second}$. Finally, nucleation of QDs occurs in the Stranski-Krastanov growth mode after deposition of nominally 0.93 mono-layer thick InAs under TMIn and AsH$_3$ flow rates of $\SI{11.8}{\micro\mole\per\minute}$ and $\SI{52.2}{\micro\mole\per\minute}$, respectively. Nucleated QDs are annealed for $\SI{60}{\second}$ at the growth temperature in AsH$_3$ ambient, before the temperature is increased for $\SI{30}{\second}$ to $\SI{515}{\degreeCelsius}$ and the annealing continues for another $\SI{30}{\second}$. Deposition of a $\SI{244}{\nano\meter}$-thick InP capping layer ($\SI{12}{\nano\meter}$ at $\SI{515}{\degreeCelsius}$, and the remaining $\SI{232}{\nano\meter}$ after increasing the temperature up to $\SI{610}{\degreeCelsius}$) finishes the growth sequence.

\subsection*{QD integration on Si substrate}\label{sec:Integration}
To integrate the QD structure on Si, a $\SI{100}{\nano\meter}$-thick layer of SiO$_2$ is deposited on top of the InP-based structure using plasma-enhanced chemical vapor deposition (PECVD) with the rate of $\SI{0.99}{\nano\meter\per\second}$, and subsequently covered by a $\SI{100}{\nano\meter}$-thick Al layer deposited via electron-beam evaporation with the rate of $\SI{90}{\angstrom\per\second}$. After flipping the structure bottom-up, it is bonded to the Si substrate. The bonding procedure includes, first, spin coating of the AP3000 adhesion promoter and benzocyclobutene (BCB) on Si and AP3000 on the InP wafer, and second, both structures are bonded at $\SI{250}{\degreeCelsius}$ in vacuum under an applied force of $\sim\SI{2}{\kilo\newton}$. Plasma ashing disposes superfluous BCB from the back-side of the InP wafer. Afterward, the InP substrate is removed in HCl and the InGaAs etch stop layer in H$_2$SO$_4$:H$_2$O$_2$:H$_2$O=1:8:80 mixture. For the sample with mesas, HSQ resist (a high purity silsesquioxane-based semiconductor grade polymer) is spin-coated and exposed using electron-beam lithography and developed in water-diluted AZ400K. The mesa pattern is then transferred to the InP by inductively coupled plasma-reactive ion etching (ICP-RIE) followed by HSQ removal in a buffered oxide etch (BHF). The mesa height is $\SI{300}{\nano\meter}$ as measured with an atomic force microscope.

\subsection*{Optical experiments}\label{sec:OptExp}
For the optical experiments, the structure is hold in a helium-flow cryostat allowing for controlled sample temperatures in the range of $\SI{4.2}{\kelvin}$ to $\SI{300}{\kelvin}$. For our standard $\upmu$PL studies, the structures are optically excited through a high numerical aperture ($\mathrm{NA}=0.4$) microscope objective with $20\times$ magnification with $\SI{660}{\nano\meter}$ or $\SI{787}{\nano\meter}$ light generated with semiconductor laser diodes, respectively. The same objective is used to collect the PL and direct it for spectral analysis to a $\SI{1}{\meter}$-focal-length monochromator equipped with a liquid-nitrogen-cooled InGaAs multichannel array detector, providing spatial and spectral resolution of $\approx \SI{2}{\micro\meter}$ and $\approx \SI{25}{\micro\electronvolt}$, respectively. Polarization properties of emitted light are analyzed by rotating a half-wave plate mounted before a fixed high-contrast-ratio ($10^6{:}1$) linear polarizer, both placed in front of the monochromator entrance.

Autocorrelation histograms, photon extraction efficiency, and TRPL are measured in a similar setup. In this setup, the structures are excited by a train of $\sim\SI{50}{\pico\second}$-long pulses with a repetition frequency of $\SI{40}{\mega\hertz}$ or $\SI{80}{\mega\hertz}$, and the central photon wavelength of $\SI{805}{\nano\meter}$. The collected photons are dispersed by a $\SI{0.32}{\meter}$-focal-length monochromator equipped either with a InGaAs multichannel array detector or NbN-based superconducting nanowire single-photon detectors (\sspds) with $\SI{\sim90}{\percent}$ quantum efficiency in the $\SI{1.5}{\micro\meter}$ to  $\SI{1.6}{\micro\meter}$ range and $\sim200$ dark counts per second. A multichannel picosecond event timer analyzes the single photon counts as time-to-amplitude converter with $\SI{256}{\pico\second}$ channel time bin width. The overall time resolution of the setup is $\SI{\sim80}{\pico\second}$.

\subsection*{Determination of the photon extraction efficiency}\label{Determination-of-PEE}

To determine the value of photon extraction efficiency $\eta_{\textrm{QD}}$, we follow the method described in Ref.~\cite{Gschrey2015}.
First, we estimate the efficiency of the setup $\eta_{\textrm{Setup}}$ by reflecting a laser tuned to the investigated QD emission range off a silver mirror placed in the setup instead of the structure.
The laser beam is attenuated with neutral density filters to achieve the \sspd~count rate in the $\si{\mega\hertz}$ range.
This number is corrected by the measured mirror reflectivity, attenuation of filters, transmission of the cryostat window and the microscope objective. Based on the laser power, the estimated setup efficiency is $\eta_{\textrm{Setup}}=(0.18~\pm~0.06)~\si{\percent}$, with the uncertainty being the standard deviation $\sigma(\eta)$ of $\eta_{\textrm{Setup}}(\lambda)$, mainly steming from the slight discrepancies in the fiber in-coupling efficiencies for different wavelengths.
Then, we excite the QDs non-resonantly with a pulsed laser diode with $f_{\textrm{rep}}=\SI{80}{\mega\hertz}$ repetition rate at the saturation power for each QD.
We collect the emission with the microscope objective ($\mathrm{NA}=0.4$), sum the \sspd~count rates for CX and X lines ($n_{\mathrm{QD}}$), as only one excitonic complex can radiatively decay at a time, and correct them by $\eta_{\textrm{Setup}}$.
Taking into account the laser repetition $f$, we estimate the photon extraction efficiency $\eta_{\textrm{QD}}=n_{\mathrm{QD}}/(f\times\eta_{\textrm{Setup}})$.
The error bars for photon extraction efficiencies are calculated by propagating the $\sigma(\eta)$ uncertainty.
This method assumes unity internal quantum efficiency of QDs ($\eta_{\textrm{int}}=\SI{100}{\percent}$), respectively the QD photon repetition rate equals $f_{\textrm{rep}}$.
It is however difficult to determine experimentally the contribution of non-radiative recombination and hence the real value of $\eta_{\textrm{int}}$. The assumption of $\eta_{\textrm{int}}=\SI{100}{\percent}$ thus determines a lower limit of $\eta_{\textrm{QD}}$ due to a possible overestimation of the total number of photons emitted by the QD ($n_{\mathrm{QD}}$).
Finally, we correct the measured $\eta$ values for the QDs A-C by the factor $\sqrt{1-g^{(2)}(0)_{\textrm{area}}}$ to account for the secondary photons due to the refilling of QD states that contribute to the measured photon flux~\cite{Yang2020,Kumano2016}.
This procedure only slightly reduces the $\eta$ values by $\SI{16}{\percent}$, $\SI{11}{\percent}$, and $\SI{5}{\percent}$ for QDs A, B, and C, respectively. With this correction, the highest $\eta$ values are $\SI{9.5}{\percent}$ and $\SI{9.1}{\percent}$ for QDs C and B, respectively.

\subsection*{Determining the single-photon purity}
For the pulsed QD excitation, we calculate the $g^{(2)}(0)$ value including the histogram background contribution $B$:
\begin{equation}
    g^{(2)}(0)_{\textrm{area}} = \int_{-T_0/2}^{T_0/2} \left[B+A\left[\exp{\left(-|\tau|/\tau_{\mathrm{dec}}\right)}-\exp{\left(-|\tau|/\tau_{\mathrm{cap}}\right)}\right]\right]\mathrm{d}\tau/
    \int_{-T_0/2}^{T_0/2} \left[B+H\exp{\left(-|\tau|/\tau_{\mathrm{dec}}\right)}\right]\mathrm{d}\tau,
\end{equation}
and with the background contribution subtracted:
\begin{equation}
    g^{(2)}(0)_{\textrm{area}} = \int_{-T_0/2}^{T_0/2} A\left[\exp{\left(-|\tau|/\tau_{\mathrm{dec}}\right)}-\exp{\left(-|\tau|/\tau_{\mathrm{cap}}\right)}\right]\mathrm{d}\tau/
    \int_{-T_0/2}^{T_0/2} H\exp{\left(-|\tau|/\tau_{\mathrm{dec}}\right)}\mathrm{d}\tau.
\end{equation}
For the histograms recorded in cw mode we use the standard equation
$
C(\tau)=N\left[1-\left(1-g^{(2)}_{\textrm{fit}}(0)\right)\exp{\left(-|\tau|/t_{\mathrm{r}}\right)}\right],
$
where $N$ is the average coincidence level at $|\tau|\gg0$.
The purity is extracted as $\mathcal{P} = 1-g^{(2)}(0)$, in particular, for the raw-data estimated purity $\mathcal{P}_{\mathrm{raw}}=C(0)/N$.

\subsection*{Numerical calculations}
\label{sec:Extraction-efficiency-calc}
The structure is modeled with a modal method (MM) employing a true open geometry boundary condition~\cite{Guer2021}.
Here, the geometry is divided into uniform layers along a propagation \textit{z} axis, and the field is expanded on eigenmodes of each uniform layer.
The eigenmode expansion coefficients in the QD layer are computed using the reciprocity theorem~\cite{Lavrinenko2014}, and the fields are connected at each layer interface using the $S$ matrix formalism~\cite{Lavrinenko2014,Li1996}.
The far field is then determined using the field equivalence principle and radiation integrals~\cite{Balanis2016}.
The extraction efficiency is defined as $\eta=P_{\textrm{lens, NA}}/P_{\textrm{in}}$, where $P_{\textrm{lens, NA}}$ is the power detected by the lens with $\textrm{NA}=0.4$ in the far field, and $P_{\textrm{in}}$ is the total power emitted from the dipole.

\section*{Acknowledgements}
We acknowledge financial support from the Danish National Research Foundation via the Research Centers of Excellence NanoPhoton (DNRF147) and the Center for Macroscopic Quantum States bigQ (DNRF142).
P.\,H. was funded by the Polish National Science Center within the Etiuda~8 scholarship (Grant~No.~2020/36/T/ST5/00511) and by the European Union under the European Social Fund.
N.\,G. acknowledges support from the European Research Council (ERC-CoG “UNITY”, Grant No. 865230), and from the Independent Research Fund Denmark (Grant No. DFF-9041-00046B).

\section*{Author contributions statement}
P. H., A. S., B. W., K. Y., and E. S. designed and fabricated the device.
P. H., A. M., A. H., and M. S. performed the experiments and analysed the data.
U. M. G., P. M., and N. G. performed the theoretical analysis of the light propagation.
P. H., A. H., M. S., and E. S. wrote the manuscript with contributions from all authors.
E.~S. initiated the research and with M. S. managed the project and supervised all efforts.

\section*{Additional information}
The authors declare no competing financial interests.
Supplementary information accompanies this paper.

\label{EndOfMain}
\clearpage
\begin{center}
\textbf{\large Supplemental Material: Bright Quantum Dot Single-Photon Emitters at Telecom Bands Heterogeneously Integrated on Si}
\end{center}
\setcounter{equation}{0}
\setcounter{figure}{0}
\setcounter{table}{0}
\setcounter{page}{1}
\setcounter{section}{0}
\renewcommand{\thesection}{S-\Roman{section}}
\makeatletter
\renewcommand{\theequation}{S\arabic{equation}}
\renewcommand{\thefigure}{S\arabic{figure}}
\renewcommand{\thetable}{S\arabic{table}}
\rfoot{\small\sffamily\bfseries\thepage/\pageref{EndOfSI}}%

\section{Identification of excitonic complexes}\label{Sec:Identification-Suppl}

	\begin{figure*}[!ht] %
		\begin{center} %
			\includegraphics[width=0.65\columnwidth]{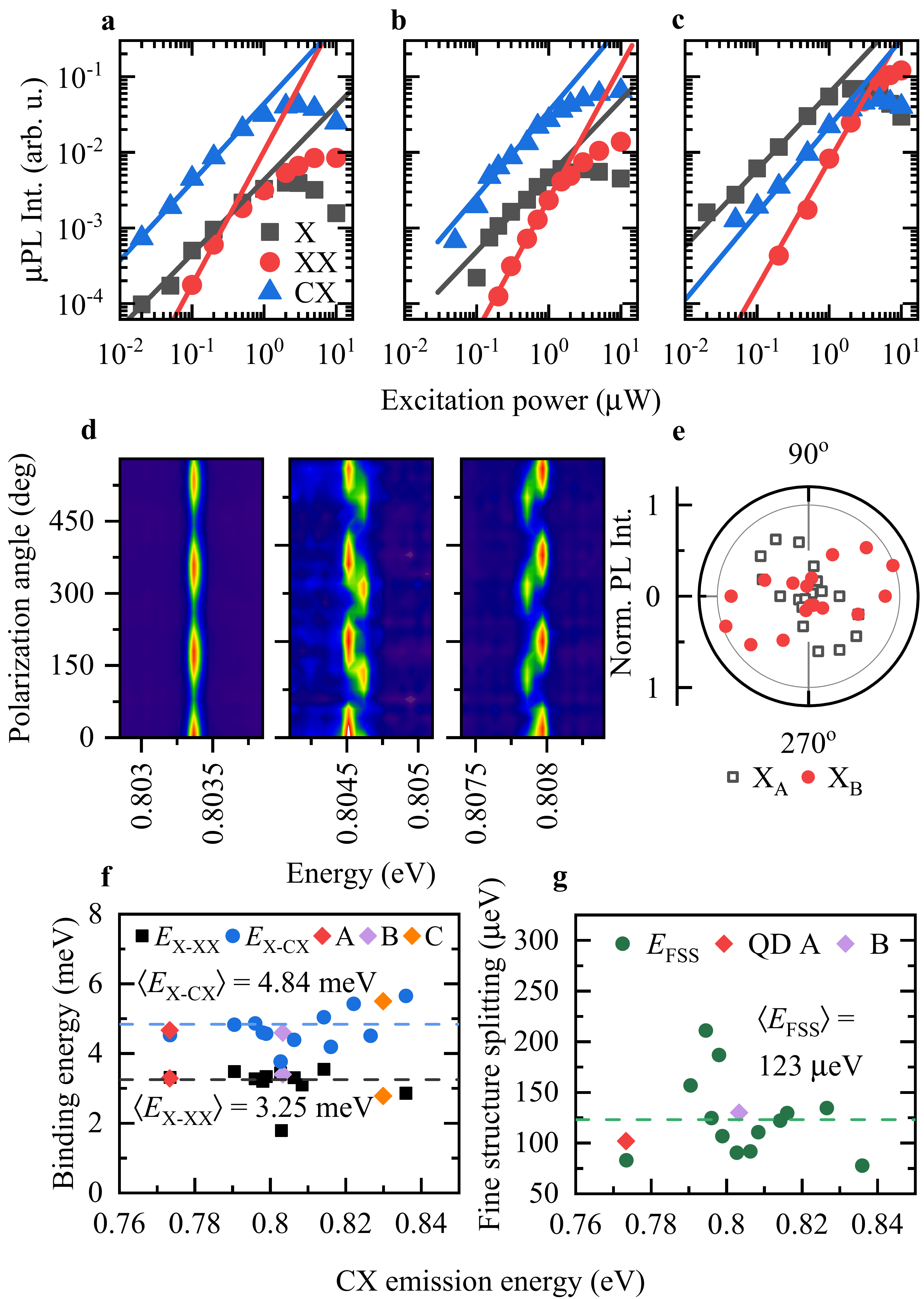} %
		\end{center} %
		\vspace{-0.8em}\caption{\label{fig:Identification-of-lines}%
			Excitonic complexes in InAs/InP QDs. 
			\textbf{a-c},~Excitation power-dependent $\upmu$PL intensity of identified lines in QDs: \textbf{a} A, \textbf{b} B, \textbf{c} C.
			\textbf{d},~Polarization-resolved $\upmu$PL signal for excitonic complexes in QD~B.
			\textbf{e},~Normalized polarization-resolved $\upmu$PL intensity for X and XX in QD~B.
			\textbf{f},~Binding energies for trions $\left( E_{\mathrm{X-CX}}\right)$ and for biexcitons $\left( E_{\mathrm{X-XX}}\right)$, with values for QDs A, B, and C marked with diamonds.
			\textbf{g},~Exciton fine structure splitting $\left(E_{\mathrm{FSS}}\right)$.}
	\end{figure*}

To demonstrate the optical properties of the QDs, three exemplary emitters representing the L- (QD~A), C- (QD~B), and S- (QD~C) telecom bands are chosen, with their $\upmu$PL spectra presented in
\subfigsref{uPL spectra}{b}{d}.
The excitonic complexes are identified based on the excitation power-dependent $\upmu$PL intensity $I_{\mathrm{\upmu PL}}$ [shown in \Supplsubfigsref{Identification-of-lines}{a}{c}], and polarization-resolved $\upmu$PL investigations [results shown in \Supplsubfigsref{Identification-of-lines}{d}{e}].
We obtain the expected~\cite{Abbarchi2009,Baier2006} linear, superlinear and almost quadratic power dependences for excitons, trions, and biexcitons, respectively, with the following exponents $b$ from fitting the power dependence $I_{\mathrm{\upmu PL}}= aP^b$ to the line intensities:
$b_{\mathrm{X}}=0.98 \pm 0.08$, $b_{\mathrm{XX}}=1.77 \pm 0.11$, $b_{\mathrm{CX}}=1.04 \pm 0.04$ (QD~A),
$b_{\mathrm{X}}=1.00 \pm 0.15$, $b_{\mathrm{XX}}=1.75 \pm 0.15$, $b_{\mathrm{CX}}=1.12 \pm 0.18$ (QD~B),
and $b_{\mathrm{X}}=1.02 \pm 0.02$, $b_{\mathrm{XX}}=1.67 \pm 0.10$, $b_{\mathrm{CX}}=1.15 \pm 0.05$ (QD~C), where $b_{\mathrm{X}}$, $b_{\mathrm{XX}}$, $b_{\mathrm{CX}}$ are exponents for the X, XX, and CX lines, respectively.

An exemplary polarization-resolved $\upmu$PL map is shown in \Supplsubfigref{Identification-of-lines}{d} with the traces that help to unambiguously ascribe the lines, with X and XX oscillating in anti-phase (right and center panel, respectively), revealing the exciton fine structure splitting with the energy $E_{\textrm{FSS}}=\SI{91}{\micro\electronvolt}$.
The slight non-orthogonality of X states, visible in the intensities of both bright exciton (X$_{\textrm{A}}$ and X$_{\textrm{B}}$) and biexciton (XX$_{\textrm{A}}$ and XX$_{\textrm{B}}$) states [\Supplsubfigref{Identification-of-lines}{e}], evidences the valence-band mixing between heavy- and light-hole states due to in-plane QD shape asymmetry and anisotropic strain effects~\cite{Leger2007,Tonin2012}.
Based on the polarization dependence of the X and CX lines, we determine the degree of linear polarization $\textrm{DOLP}= (39.6 \pm 3.2)~\si{\percent}$ and the amplitude of the hole states mixing $\beta=(33.7 \pm 2.7)~\si{\percent}$~\cite{Tonin2012}.
For the CX, we observe the lack of the emission energy dependence on the linear polarization angle [\Supplsubfigref{Identification-of-lines}{d}, left panel].

We perform in an similar manner the identification of complexes for other QDs in the investigated structures.
Based on polarization-resolved and excitation power-dependent $\upmu$PL spectra of InAs/InP QDs in the structure with mirror which emit in the range of $\SI{1.48}{\micro\meter}$ to $\SI{1.6}{\micro\meter}$, we determine the binding energies for trions $\left( E_{\mathrm{X-CX}}\right)$ and for biexcitons $\left( E_{\mathrm{X-XX}}\right)$.
The summary of determined binding energies for biexcitons and trions, and the $E_{\textrm{FSS}}$ values are shown in~\Supplsubfigsref{Identification-of-lines}{f}{g}.
We find that the distributions of biexciton and trion binding energies are narrow and the values for QDs A-C are close to their typical values (see the diamonds marking values for QDs A, B, and C).
We calculate the average values of $\langle E_{\mathrm{X-CX}} \rangle = \SI{4.84}{\milli\electronvolt}$ and $\langle E_{\mathrm{X-XX}} \rangle = \SI{3.25}{\milli\electronvolt}$.
The obtained values are spectrally-independent and only slightly spread around the average values, therefore they help in the identification of excitonic complexes in the investigated QDs A-C.
For the fine structure splitting energy, we calculate the average value of $\langle E_{\mathrm{FSS}} \rangle = \SI{123}{\micro\electronvolt}$.
	\begin{figure*}[!ht] %
		\begin{center} %
			\includegraphics[width=0.75\columnwidth]{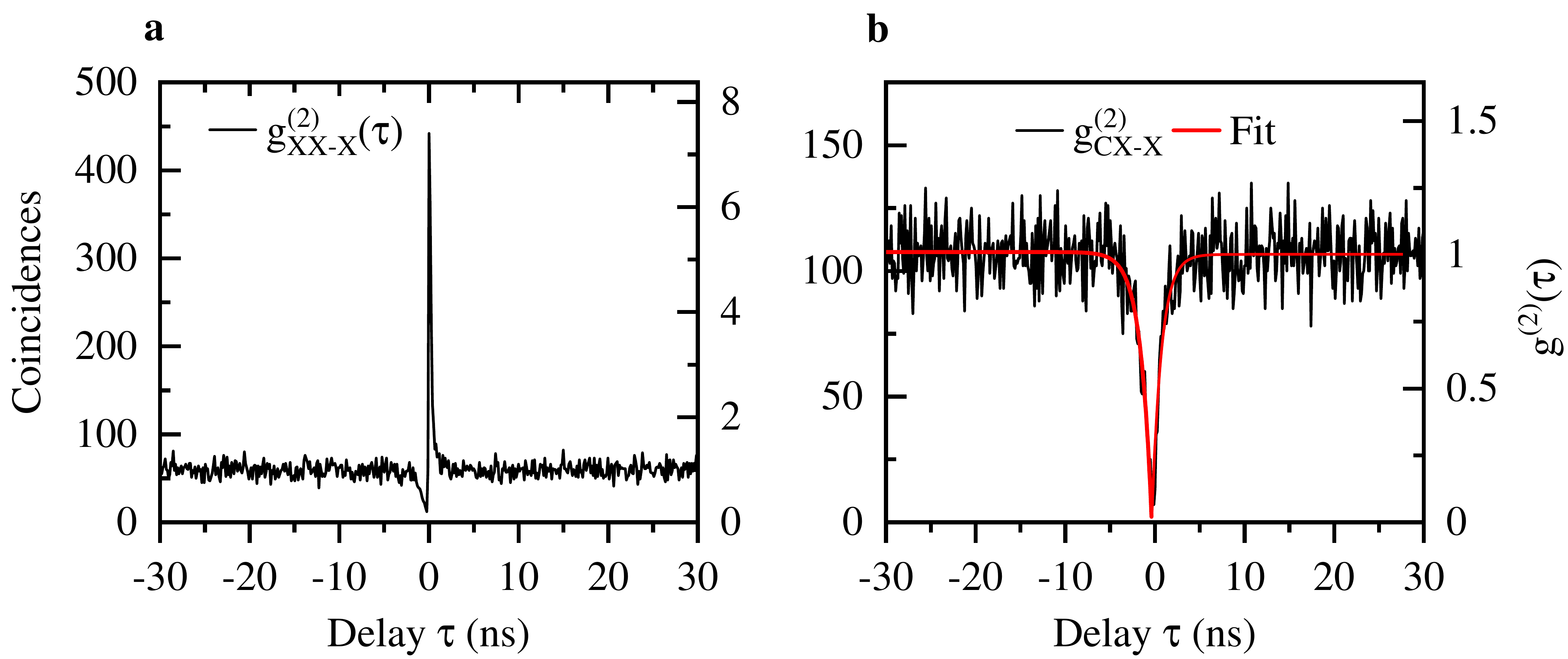} %
		\end{center} %
		\vspace{-0.8em}\caption{\label{fig:Suppl-Cross-correlation}%
			\textbf{a},~Cross-correlation of XX and X lines of QD~C.
			\textbf{b},~Cross-correlation of CX and X lines of QD~C with fit line (red).}
	\end{figure*}

Finally, for QD~C we present the cross-correlation between XX-X and CX-X lines to unambiguously prove the identification of excitonic lines and the fact that they origin from the same QD.
Here, X emission events are registered by the stopping detector.
The histograms are shown in \Supplsubfigref{Identification-of-lines}{g} for XX-X and \Supplsubfigref{Identification-of-lines}{h} for CX-X cross-correlation measurements.
In the case of XX-X, we find a strong bunching for positive delays, evidencing the cascaded XX-X emission, while for the CX-X case we see an asymmetric dip which we fit with the function of the form 
$g^{(2)}_{\mathrm{CX-X}}(\tau) = A\left[1-\exp{\left(\tau/t_{\mathrm{r}}\right)}\right]$ separately for $\tau<0$ and $\tau>0$, where $A$ is a scaling factor, $\tau$ is the time delay, and $t_{\mathrm{r}}$ is the antibunching time constant.
We find that the $t_{\mathrm{r}}$ is different for positive and negative delays: $t_{\mathrm{r}}=1.25 \pm 0.08~\si{\nano\second}$ and $t_{\mathrm{r}}=1.06 \pm 0.08~\si{\nano\second}$, respectively, corresponding with the CX and X emission.

\section{FDTD calculations}\label{Sec:FDTD-calculations}
\begin{figure*}[!ht] %
		\begin{center} %
			\includegraphics[width=0.75\columnwidth]{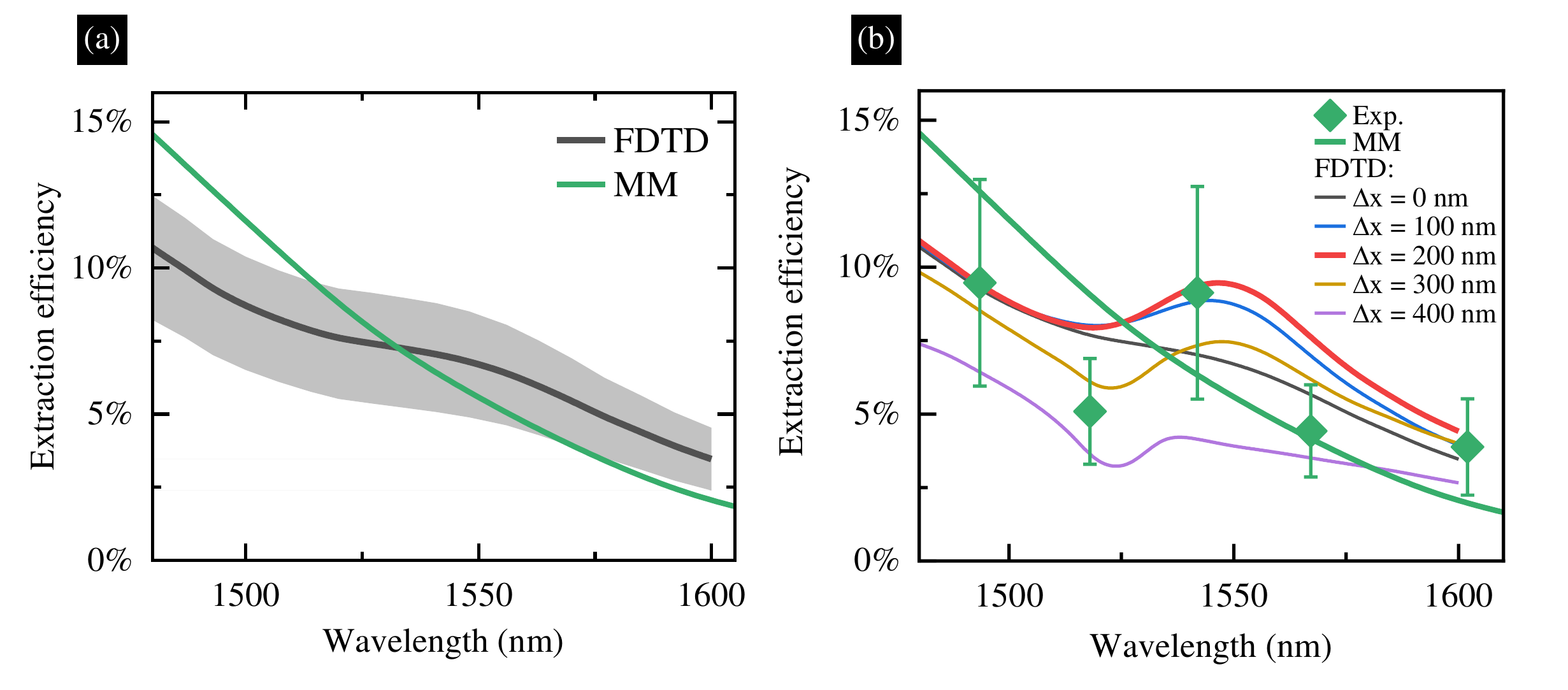} %
		\end{center} %
		\vspace{-0.8em}\caption{\label{fig:FDTD-calculations}%
		Analysis of the influence of the QD position in the mesa on the extraction efficiency.
		\textbf{a},~Comparison of calculated extraction efficiency between the FDTD and MM methods for the QD in the center of the mesa ($\Delta x=0~\si{\nano\meter}$). The shaded uncertainty range results from different vertical screen positions ($\SI{0.8}{\micro\meter}$ to $\SI{2.4}{\micro\meter}$ above the mesa surface)
    	\textbf{b},~Dispersion of the extraction efficiency for different QD displacement from the center of the mesa ($\Delta x$).
}
	\end{figure*}

In this section we investigate the influence of the QD position displacement in the mesa structure on the extraction efficiency.
For that purpose, we employ finite-difference time-domain (FDTD) 3D Electromagnetic Simulator provided by Lumerical Inc.~\cite{Lumerical}, as a complementary tool to the previous one based on the modal method (MM)~\cite{Guer2021}.
More details of the employed FDTD method can be found elsewhere~\cite{Mrowinski2019}.
In order to establish convergence between the two numerical methods we first compare the results for the identical photonic mesa structures with $\mathrm{D}=\SI{2}{\micro\meter}$ containing a point dipole at the central position $\Delta x=0~\si{\nano\meter}$.
Then, the FDTD approach was tuned to minimize deviation with respect to the modal method by slight change of the numerical aperture of the collected emission or the position of the 2D field-power monitor located above the mesa structure. Such tuning mechanism is visualized in \Supplsubfigref{FDTD-calculations}{a} as the black shaded area which overall is qualitatively similar to the results obtained by the modal method. 

Next, such consistent FDTD model was used to calculate the dispersion of extraction efficiency for a variety of the dipole displacements in the range of $\Delta x=\SI{0}{\nano\meter}$ to $\SI{400}{\nano\meter}$.
Starting from $\Delta x=\SI{100}{\nano\meter}$, a local enhancement of extraction efficiency around $\SI{1.55}{\micro\meter}$ is already observable and this effect is further strengthened for $\Delta x=\SI{200}{\nano\meter}$ displacement, as it is shown in \Supplsubfigref{FDTD-calculations}{b}.
Further shift of the dipole position causes decrease of the extraction efficiency values, and taking into account the experimental results we observe the best match for QD B at $\Delta x=\SI{200}{\nano\meter}$, suggesting that the QDs are placed in between $\Delta x\approx\SI{0}{\nano\meter}$ to $\SI{200}{\nano\meter}$.

\section{Time-resolved microphotoluminescence for CX lines in QDs A-C}
	\begin{figure*}[!ht] %
		\begin{center} %
			\includegraphics[width=\columnwidth]{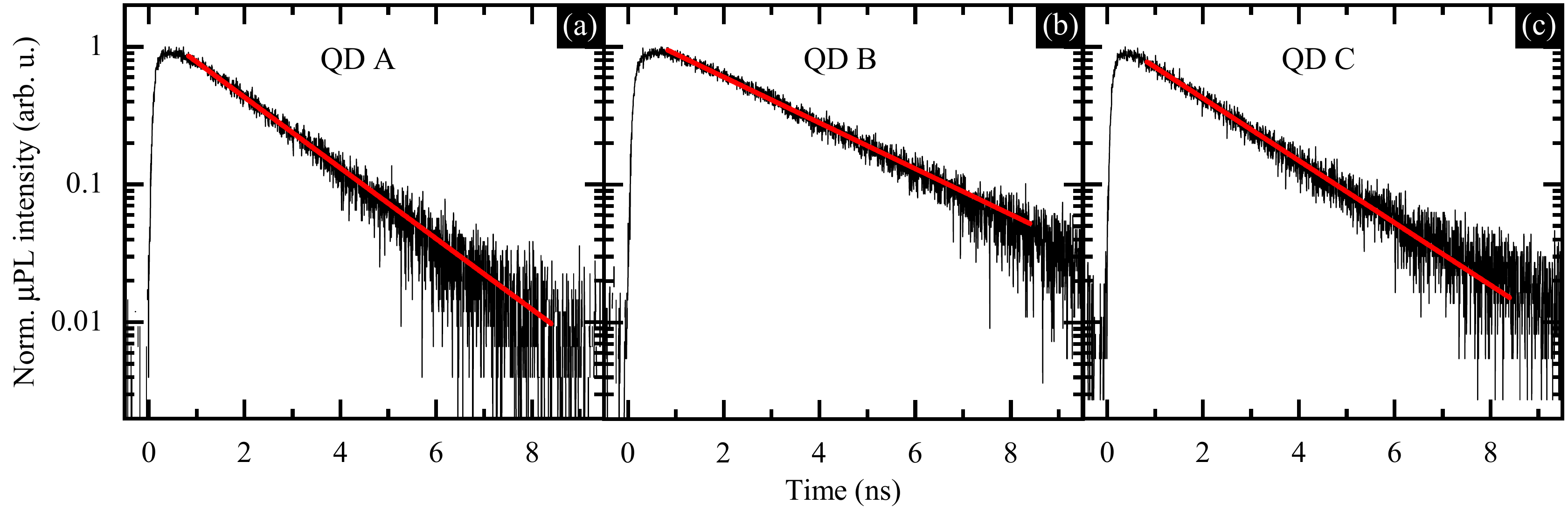} %
		\end{center} %
		\vspace{-0.8em}\caption{\label{fig:TRPL}%
		Time-resolved $\upmu$PL traces for trions in QDs A-C: \textbf{a},~QD~A, \textbf{b},~QD~B, \textbf{c},~QD~C.
		Red solid lines are fit lines to the experimental data (black lines), according to Eq.~\eqref{Eq:TRPL}.}
	\end{figure*}

Low-temperature ($T=\SI{4.2}{\kelvin}$) time-resolved $\upmu$PL (TRPL) traces registered for QDs A, B, and C are presented in \Supplfigref{TRPL} with solid black lines.
Each trace is best fitted with a single-exponential decay function (red solid lines) to extract decay time constant $\tau_{\textrm{PL}}$.
We use the function of the form:
\begin{equation}\label{Eq:TRPL}
\centering
	I(t) = A\exp{\left(-t/\tau_{\textrm{PL}}\right)},
\end{equation}
where $I(t)$ is the TRPL intensity at time $t$, and $A$ is the amplitude of the signal.
The extracted $\tau_{\textrm{PL}}$ values are 
$1.69 \pm 0.01~\si{\nano\second}$ (QD~A),
$2.61 \pm 0.01~\si{\nano\second}$ (QD~B), and
$1.92 \pm 0.01~\si{\nano\second}$ (QD~C).
The $\tau_{\mathrm{PL}}$ times are rather typical for single InAs/InP QDs, independently of their exact size and symmetry~\cite{Takemoto2007,Dusanowski2018,Syperek2013APL,Musial2019} and agree with the $\tau_{\mathrm{dec}}$ values (see comparison in \Suppltabref{g2-pulsed}).

\section{CW histograms}\label{Sec:g2-Supplemental}

	\begin{figure*}[!ht] %
		\begin{center} %
			\includegraphics[width=\columnwidth]{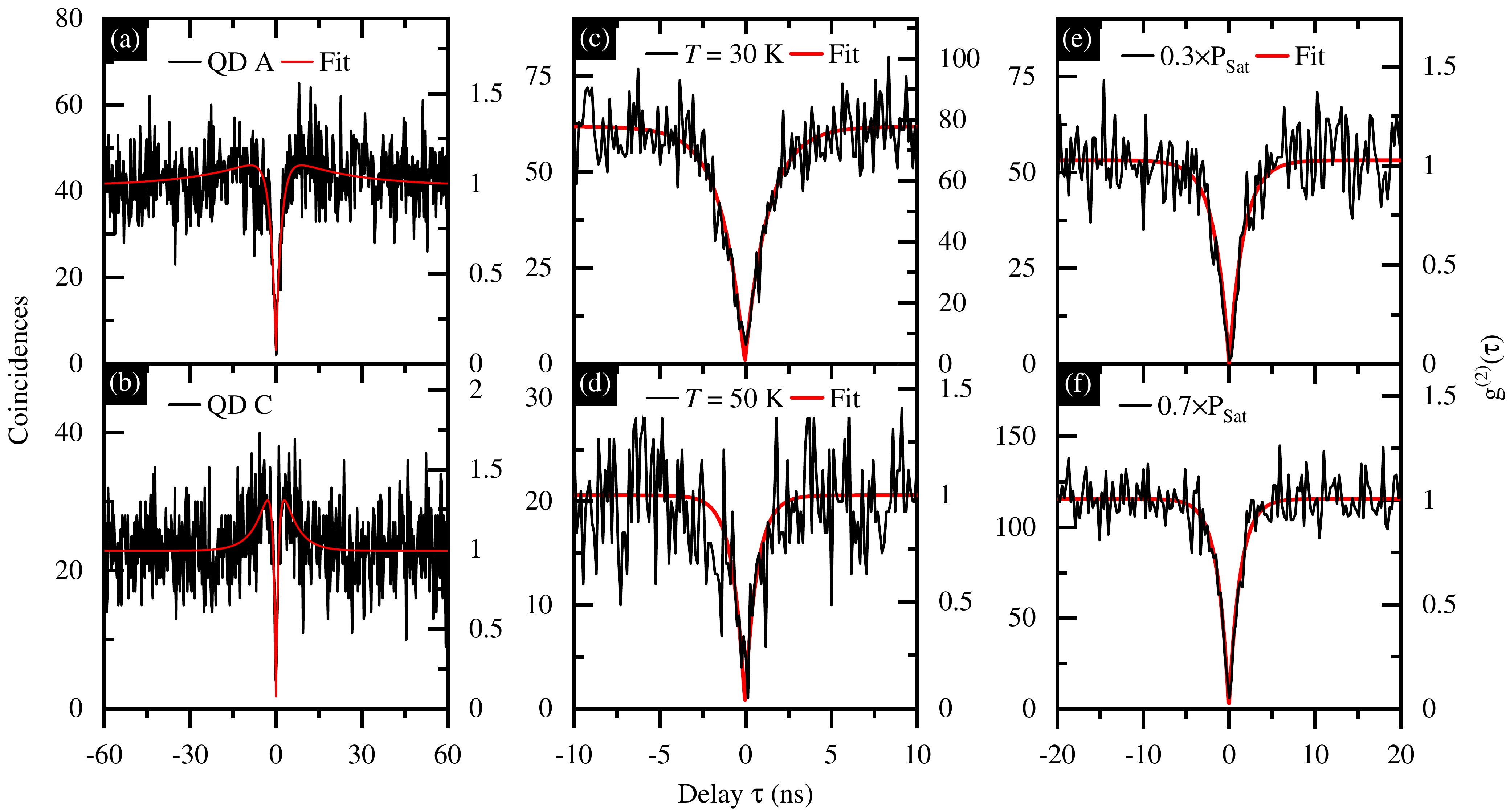} %
		\end{center} %
		\vspace{-0.8em}\caption{\label{fig:g2-Supplemental}%
Evaluation of the single-photon emission purity for CX lines under cw excitation in QD \textbf{a},~A and \textbf{b},~C, and for CX line in QD~B at elevated temperatures: recorded and C QD~B, at \textbf{c},~$T=\SI{30}{\kelvin}$, \textbf{d},~$T=\SI{50}{\kelvin}$, and for \textbf{e},~$0.3\times P_{\textrm{Sat}}$ and \textbf{f},~$0.7\times P_{\textrm{Sat}}$.}
	\end{figure*}

In this section we present the histograms that broaden the discussion of the single-photon emission quality for investigated SPEs, presented in \figref{g2-pulsed} in the main article.
Here, we focus on the cw excitation of the CX lines, and we show the  autocorrelation histograms for QDs A and C in \Supplsubfigref{g2-Supplemental}{a} and \Supplsubfigref{g2-Supplemental}{b}, respectively, together with fit lines.
The histograms were recorded for the laser excitation power corresponding to the $0.7\times P_{\textrm{Sat}}$ of the respective lines.
We observe an additional weak bunching effect which we attribute to the blinking caused by the interaction with carrier traps in the QD vicinity or background doping~\cite{Dalgarno2008,Kumano2016}, and we fit the normalized histograms with the function~\cite{Benyoucef2013}
\begin{equation}
\centering
  g^{(2)}(\tau)=1-A\exp{\left(-|\tau|/\tau_1\right)}+B\exp{\left(-|\tau|/\tau_2\right)},
\end{equation}
where A and B are fit parameters, while $\tau_1$ and $\tau_2$ are antibunching and bunching time constants, respectively.
The $g^{(2)}(0)$ value is obtained as $g^{(2)}(0)=1-A+B$.
Determined parameters are 
$g^{(2)}(0)=0.074$ ($\sigma=0.062$),
$\tau_1=1.93 \pm 0.15~\si{\nano\second}$, and
$\tau_2=22.7 \pm 3.0~\si{\nano\second}$ for QD~A, and
$g^{(2)}(0)=0.07$ ($\sigma=0.11$),
$\tau_1=0.97 \pm 0.09~\si{\nano\second}$, and
$\tau_2=4.81 \pm 0.45~\si{\nano\second}$ for QD~C.
These $g^{(2)}(0)$ values are displayed in \Suppltabref{g2-pulsed-extr-params} for their easier comparison.

Then, we show the histograms for CX in QD~B registered under the same excitation conditions as in \subfigsref{g2-pulsed}{b}{c}, except for the temperature of the structure (\Supplsubfigsref{g2-Supplemental}{c}{d}) and excitation power (\Supplsubfigsref{g2-Supplemental}{e}{f}).
We fit the normalized histograms with the standard function (see Methods).
For the Stirling-compatible temperatures of $T=\SI{30}{\kelvin}$ and $T=\SI{50}{\kelvin}$ we obtain almost perfect single-photon emission with the purity of $g^{(2)}(0)_{\SI{30}{\kelvin}}=0$ ($\sigma=0.054$, \Supplsubfigref{g2-Supplemental}{c}) and $g^{(2)}(0)_{\SI{50}{\kelvin}}=0.017$ ($\sigma=0.096$, \Supplsubfigref{g2-Supplemental}{d}).

Next, for the C-band QD~B we study the quantum nature of the emission in the excitation power-dependent photon autocorrelation measurements under cw non-resonant excitation.
\Supplsubfigsref{g2-Supplemental}{e}{f} present the autocorrelation histograms corresponding to the excitation of the CX line under $0.3\times P_{\textrm{Sat}}$ [\Supplsubfigref{g2-Supplemental}{e}], and $0.7\times P_{\textrm{Sat}}$ [\Supplsubfigref{g2-Supplemental}{f}], where $P_{\textrm{Sat}}$ is the laser excitation power corresponding to the saturation of the line's $\upmu$PL intensity.
We fit the histograms (see Methods), and achieve the values of $g^{(2)}(0)=0$ for all probed excitation powers, with $\sigma=$ 0.075, 0.056, 0.038 for $0.3\times P_{\textrm{Sat}}$, $0.7\times P_{\textrm{Sat}}$, and $P_{\textrm{Sat}}$\footnote{The histogram recorded for $P_{\textrm{Sat}}$ is shown in \subfigref{g2-pulsed}{b}}, respectively.
The increasing pump rate $W_p$ results in the decrease of $t_{\mathrm{r}}$ so that $t_{\mathrm{r}}=(1.70 \pm 0.19)~\si{\nano\second}$, $t_{\mathrm{r}}=(1.31 \pm 0.11)~\si{\nano\second}$, and $t_{\mathrm{r}}=(1.12 \pm 0.06)~\si{\nano\second}$ for $0.3\times P_{\textrm{Sat}}$, $0.7\times P_{\textrm{Sat}}$, and $P_{\textrm{Sat}}$, respectively.	

The obtained $g^{(2)}(0)$ for CX in the QD B are repeated in \Suppltabref{Summary-fitted-parameters-cw}.

\section{Temperature-dependent photoluminescence of QD B}\label{Sec:QD-B-Temp-analysis}

\begin{figure*}[!ht] %
		\begin{center} %
			\includegraphics[width=\columnwidth]{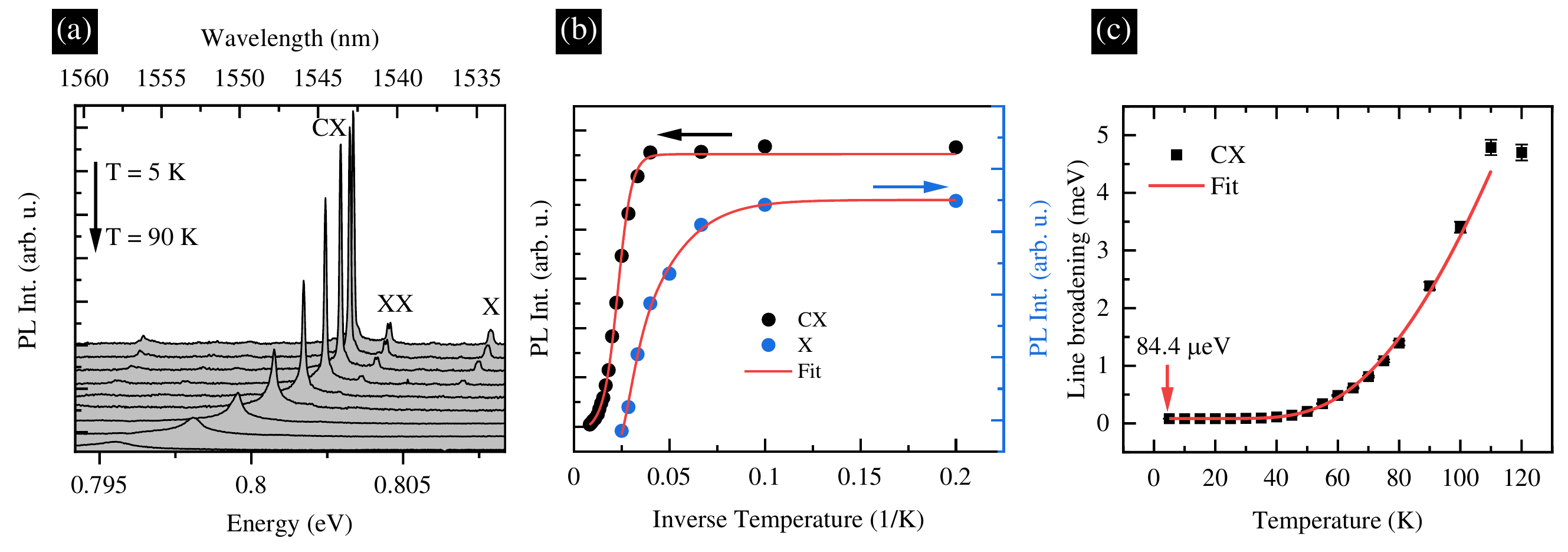} %
		\end{center} %
		\vspace{-0.8em}\caption{\label{fig:QD-B-Temp-analysis}%
		Analysis of the temperature-dependent $\upmu$PL spectra for QD B.
		\textbf{a},~$\upmu$PL spectra taken at different temperatures in the range of $T = \SI{5}{\kelvin}$ to $\SI{90}{\kelvin}$.
		\textbf{b},~$\upmu$PL intensity for X and CX lines fitted with the Arrhenius formula (red lines), according to Eq.~\eqref{ArrhEq}.
		\textbf{c},~Broadening of the CX line with the fitted temperature dependence of Eq.~\eqref{Moody}.}
	\end{figure*}

The temperature-dependent $\upmu$PL of the QD B is recorded in the temperature range of $T = \SI{5}{\kelvin}$ to $\SI{120}{\kelvin}$ and analyzed before recording the autocorrelation histograms at elevated temperatures, shown in \subfigref{g2-pulsed}{c}.
The spectra, the temperature-dependent quench of CX and X lines, and the linewidth broadening of the CX line are presented in \Supplfigref{QD-B-Temp-analysis}.

To identify the most efficient carrier excitation channels, the temperature-dependent $\upmu$PL intensity is fitted with a standard formula assuming two activation processes~\cite{LambkinAPL1990}:	%
	\begin{equation}\label{ArrhEq}
	\centering
		I\left( T \right) = \frac{I_0}{1+B_{\mathrm{1}} \exp{\left(-E_{\mathrm{a,1}}/k_{\mathrm{B}}T\right)}+
		B_{\mathrm{2}} \exp\left( -E_{\mathrm{a,2}}/k_{\mathrm{B}}T\right)},
	\end{equation}
where $I_0$ is the PL intensity for $T\to 0$, $E_{\mathrm{a,1}}$ and $E_{\mathrm{a,2}}$ are activation energies, and $B_{\mathrm{1}}$ and $B_{\mathrm{2}}$ are relative rates corresponding to the efficiency of involved processes.
We achieve the activation energies of 
$E_{\mathrm{a,1}} = 23.6 \pm 0.8~\si{\milli\electronvolt}$ and $E_{\mathrm{a,2}} = 81.1 \pm 15~\si{\milli\electronvolt}$ for CX and
$E_{\mathrm{a,1}} = 5.0 \pm 1.5~\si{\milli\electronvolt}$ and $E_{\mathrm{a,2}} = 22.2 \pm 7.3~\si{\milli\electronvolt}$ for X.
We find that for both lines the dominant process responsible for the $\upmu$PL intensity quench is the one corresponding to the $E_{\mathrm{a,2}}$ energy, as the rates ratio $B_{\mathrm{1}}:B_{\mathrm{2}}$ is $1:264$ and $1:402$ for CX and X lines, respectively.
Interestingly, both complexes share the activation energy in the range of $\SI{\sim22}{\milli\electronvolt}$ to $\SI{24}{\milli\electronvolt}$, in agreement with the value obtained for the CX line in similar QDs~\cite{Holewa2020PRB}, where the values of $E_{\mathrm{a,1}} = 0.8 \pm 0.5 \si{\milli\electronvolt}$ and $E_{\mathrm{a,2}} = 23 \pm 5 \si{\milli\electronvolt}$ were reported for a CX.
There, the process corresponding to the $E_{\mathrm{a,2}}$ activation energy was attributed as the charge transfer to higher orbital states, based on the band structure calculations within the 8-band $k\cdot p$ framework.
Furthermore, for a QD emitting around the $\SI{1.55}{\micro\meter}$ spectral range, the calculated energy distance for holes between their QD and WL ground states is in the range of $\SI{70}{\milli\electronvolt}$ to $\SI{90}{\milli\electronvolt}$~\cite{Holewa2020PRB}, therefore we attribute the $E_{\mathrm{a,2}} = 81.1 \pm 15 \si{\milli\electronvolt}$ estimated for the CX as the excitation of the hole from the QD B to the WL ground state.

The temperature dependence of linewidth was fitted with the formula that includes the contribution of thermally-activated phonon sidebands to the zero-phonon line~\cite{Gammon1996Science,Moody2011},
\begin{equation}\label{Moody}
\centering
	\varGamma(T)
	= \varGamma \left(\SI{4.2}{\kelvin}\right) + a\,\left[ \exp\left( \frac{E_{\mathrm{ph,\varGamma}}}{k_{\mathrm{B}}T} \right) - 1 \right]^{-1},
	\end{equation}
where $k_{\mathrm{B}}$ is the Boltzmann constant, parameter $a=72.8 \pm 6.8 ~\si{\milli\electronvolt}$ and $E_{\mathrm{ph,\varGamma}}=27.39\pm0.55~\si{\milli\electronvolt}$ is an average energy of phonons.
From the fitting we obtain the initial CX line broadening of $\varGamma\left(\SI{4.2}{\kelvin}\right)=84.4 \pm 1.2~\si{\micro\electronvolt}$ what is above the lifetime-limited linewidth ($\SI{0.5}{\micro\electronvolt}$ to $\SI{0.8}{\micro\electronvolt}$, according to TRPL data in \Supplfigref{TRPL}) as well as above the spectral resolution of the $\upmu$PL setup ($\SI{\sim25}{\micro\electronvolt}$) and so points to the presence of spectral diffusion due to the deep charge traps in the QD vicinity, also on the etched mesa walls~\cite{Ortner2004}.
On the other hand, the initial broadening is much lower than in the previously fabricated structure with InAs/InP QDs in mesas of $\SI{340}{\micro\electronvolt}$~\cite{Holewa2020PRB}.

\section{Summary of fitted and derived parameters}\label{Sec:Summary-fitted-parameters}
In \Suppltabref{g2-cw} we provide the fit parameters for the histograms obtained under cw laser excitation for QDs A, B, and C.
The purity $\mathcal{P}$ is defined as $\mathcal{P}=1-g^{(2)}(0)$.
For the purity determination we take the fitted values of $g^{(2)}(0)_{\textrm{fit}}$ unless $g^{(2)}(0)_{\textrm{fit}}=0$.
In such cases, we employ the more conservative estimation of purity, utilizing the $g^{(2)}(0)_{\textrm{raw}} = C(0)/N$ value (see Methods for the fitting formula).

		\begin{table*}[ht!]
		\centering
		\begin{threeparttable}
		\begin{tabular}{|c|c|c|c|}
		\hline
 	\multicolumn{4}{|c|}{Different QDs, cw excitation}\\\hline
 	QD & $g^{(2)}(0)_{\textrm{fit}}$ & $g^{(2)}(0)_{\textrm{raw}}$ & $\mathcal{P}$ \\\hline
 	A & 0 ($\sigma=0.11$) & 0.0480 & $\left(95.2^{+4.8}_{-6.2}\right)\%$ \\
 	B & 0 ($\sigma=0.056$)\tnote{*} & $0.0519$ & $\left(94.8^{+5.2}_{-0.4}\right)\%$ \\
 	C & 0.07 ($\sigma=0.11$) & 0.173 & $\left(93^{+7}_{-11}\right)\%$ \\
 				\hline\hline
		\end{tabular}\vspace{-0.7em}
		\vspace{1em}
		  \begin{tablenotes}
  \item[*] At $0.7\times P_{\mathrm{Sat}}$.
  \end{tablenotes}
		\vspace{-0.4em}\caption{\label{tab:g2-cw} Fit parameters of single-photon emission under cw excitation (\Supplsubfigsref{g2-Supplemental}{a}{b} for QDs A and C, \Supplsubfigref{g2-Supplemental}{f} for QD B).}
	\end{threeparttable}
	\end{table*}

In \Suppltabref{g2-pulsed} we give the fit parameters for histograms obtained for the pulsed laser excitation (see Methods for the fitting formula).
$B$ is the level of background coincidences, $A$ is a scaling parameter related to secondary photon emission, $H$ is an average non-zero peak height, $\tau_{\mathrm{dec}}$ and $\tau_{\mathrm{cap}}$ are the decay and capture time constants, respectively.
The corresponding histograms are shown in \figref{g2-pulsed} and \Supplsubfigsref{g2-Supplemental}{a}{b}.
We give also the $\upmu$PL decay time $\tau_{\mathrm{PL}}$ recorded in time-resolved $\upmu$PL experiment (the corresponding $\upmu$PL decay traces are presented in \Supplfigref{TRPL}).
The uncertainties given in \Suppltabref{g2-pulsed} are standard errors of the fitting procedure ($\sigma$).

		\begin{table*}[ht!]
		\centering
		\begin{threeparttable}
		\begin{tabular}{|c|c|c|c|c|c|c|}
		\hline
			& 
			\multicolumn{5}{|c|}{Different QDs, pulsed excitation} &
			TRPL\\ \hline
			QD &
			$B$ &
			$A$ &
			$H$ &
			$\tau_{\mathrm{cap}}$ &
			$\tau_{\mathrm{dec}}$ &
			$\tau_{\mathrm{PL}}$\\ \hline
			A &   $0.86 \pm 0.05$ & $12.49 \pm 0.90$ & $37.30 \pm 0.16$ & $0.34 \pm 0.07~\si{\nano\second}$ & $1.91 \pm 0.02~\si{\nano\second}$ & $1.69 \pm 0.01~\si{\nano\second}$\\
			B &  $13.32 \pm 0.16$ & $176 \pm 15$ & $152.35 \pm 0.47$ & $2.28 \pm 0.05~\si{\nano\second}$ & $2.80 \pm 0.02~\si{\nano\second}$ & $2.61 \pm 0.01~\si{\nano\second}$\\
			C &  $1.41 \pm 0.15$ & $11.2 \pm 2.0$ & $76.82 \pm 0.58$ & $0.44 \pm 0.17~\si{\nano\second}$ & $1.99 \pm 0.02~\si{\nano\second}$ & $1.92 \pm 0.01~\si{\nano\second}$\\
			\hline\hline
		\end{tabular}\vspace{-0.7em}
		\vspace{1em}
		  \begin{tablenotes}
  \item[*] At $0.7\times P_{\mathrm{Sat}}$.
  \end{tablenotes}
		\vspace{-0.4em}\caption{\label{tab:g2-pulsed} Fit parameters of single-photon emission under and pulsed excitation for QDs A-C (\subfigref{g2-pulsed}{a}) with PL decay times (\figref{TRPL}).}
	\end{threeparttable}
	\end{table*}

\newpage
\Suppltabref{g2-pulsed-extr-params} gives the derived $g^{(2)}(0)$ function values for pulsed excitation together with corresponding purity. 
We give both the $g^{(2)}(0)$ value based on the level of coincidences at $\tau=0$ compared with $H$, and based on the area under the zero histogram peak.
See Methods for the applied formulas with and without the background correction.
The uncertainties given in \Suppltabref{g2-pulsed-extr-params} are combined standard uncertainties based on $A$ and $H$ standard fitting errors.

		\begin{table*}[ht!]
		\centering
		\begin{threeparttable}
		\begin{tabular}{|c|c|c|c|c|c|c|}
		\hline
			\multicolumn{7}{|c|}{Different QDs, pulsed excitation}\\ \hline
			& \multicolumn{2}{|c|}{Function value approach} & 
			\multicolumn{4}{|c|}{Peak area approach}\\ \hline
			& \multicolumn{2}{|c|}{Fit value at $\tau=0$} &
			\multicolumn{2}{|c|}{Histogram background included} &
			\multicolumn{2}{|c|}{Histogram background subtracted}\\ \hline
			QD &
			$g^{(2)}(0) = B/H$ &
			$\mathcal{P} = 1-g^{(2)}(0)$ &
			$g^{(2)}(0)_{\textrm{area}}$ &
			$\mathcal{P}_{\mathrm{area}}$ &
			$g^{(2)}(0)_{\textrm{area}}$ & $\mathcal{P}_{\mathrm{area}}$\\ \hline
			A & $0.023 \pm 0.010$ & $97.7 \pm 1.0~\si{\percent}$ & $0.371 \pm 0.020$ & $62.9 \pm 2.0~\si{\percent}$ & $0.276 \pm 0.002$ & $72.4 \pm 2.0~\si{\percent}$ \\
			B & $0.087 \pm 0.017$ & $91.3 \pm 1.7~\si{\percent}$ & $0.433 \pm 0.018$ & $56.7 \pm 1.8~\si{\percent}$ & $0.209 \pm 0.018$ & $79.1 \pm 1.8~\si{\percent}$\\
			C & $0.018 \pm 0.012$ & $98.2 \pm 1.2~\si{\percent}$ & $0.205 \pm 0.02$ & $79.5 \pm 2.0~\si{\percent}$ & $0.114 \pm 0.02$ & $88.6 \pm 2.0~\si{\percent}$\\
			\hline\hline
		\end{tabular}\vspace{-0.7em}
		\vspace{1em}
		  \begin{tablenotes}
  \item[*] At $0.7\times P_{\mathrm{Sat}}$.
  \end{tablenotes}
		\vspace{-0.4em}\caption{\label{tab:g2-pulsed-extr-params} $g^{(2)}(0)$ values and corresponding purity $\mathcal{P}$ of the single-photon emission under pulsed excitation (\subfigref{g2-pulsed}{a}) derived based on the fit parameters given in \Suppltabref{g2-pulsed}.}
	\end{threeparttable}
	\end{table*}
\Suppltabref{Summary-fitted-parameters-cw} displays the fit parameters obtained for CX in QD B, for the excitation-power- and temperature-dependent autocorrelation histograms.
Again, for the determination of $\mathcal{P}$ we use the fitted values of $g^{(2)}(0)_{\textrm{fit}}$ unless $g^{(2)}(0)_{\textrm{fit}}=0$.
In such cases, we utilize the $g^{(2)}(0)_{\textrm{raw}} = C(0)/N$ value.

		\begin{table*}[ht!]
		\centering
		\begin{threeparttable}
		\begin{tabular}{|c|c|c|c|c|c|c|c|}
		\hline
		    \multicolumn{8}{|c|}{QD B, cw excitation}\\\hline
		    \multicolumn{4}{|c|}{Excitation power series} & \multicolumn{4}{|c|}{Temperature series} \\\hline
		    Laser power & $g^{(2)}(0)_{\textrm{fit}}$ & $g^{(2)}(0)_{\textrm{raw}}$ & $\mathcal{P}$ & Temperature & $g^{(2)}(0)_{\textrm{fit}}$ & $g^{(2)}(0)_{\textrm{raw}}$ & $\mathcal{P}$\\ \hline
            $0.3\times P_{\textrm{Sat}}$ & 0 ($\sigma=0.075$) & $0.0188$ & $\left(98.1^{+1.9}_{-5.6}\right)\%$ &
            $\SI{30}{\kelvin}$ & 0 ($\sigma=0.054$) & $0.0808$ & $91.9 \pm 5.4\%$ \\
            $0.7\times P_{\textrm{Sat}}$ & 0 ($\sigma=0.056$) & $0.0519$ & $\left(94.8^{+5.2}_{-0.4}\right)\%$ &
            $\SI{50}{\kelvin}$ & 0.017 ($\sigma=0.096$) & 0.0486 & $\left(98.3^{+1.7}_{-7.9}\right)\%$ \\
            $P_{\textrm{Sat}}$ & 0 ($\sigma=0.038$) & $0.0272$ & $\left(97.3^{+2.7}_{-1.1}\%\right)$ &
            $\SI{80}{\kelvin}$ & 0.25 ($\sigma=0.19$) & 0.299 & $75\pm19\%$\\
			\hline\hline
		\end{tabular}\vspace{-0.7em}
		\caption{\label{tab:Summary-fitted-parameters-cw} \label{EndOfSI} The histogram fit parameters obtained for CX in QD B, for excitation power- and temperature-dependent autocorrelation histograms [see \subfigsref{g2-pulsed}{b}{c} and \Supplsubfigsref{g2-Supplemental}{c}{f}].
		}
	\end{threeparttable}
	\end{table*}

\FloatBarrier
\end{document}